\documentclass[twocolumn,showpacs,amsfonts,amssymb,amsmath,floats,superscriptaddress,aps]{revtex4-1}
\usepackage{graphicx}
\usepackage{dcolumn}
\usepackage{multirow}
\usepackage{color}
\usepackage{caption,subcaption}

\usepackage{ulem} 

\usepackage[outdir=./]{epstopdf}
\usepackage{upgreek}

\def\expect#1{\left\langle#1\right\rangle}
\def\bra#1{\langle#1|}
\def\ket#1{|#1 \rangle}

\def\w{\omega}

\def\Tr#1{{\rm Tr}\left[#1\right]}
\def\ul#1{\underline{#1}}
\def\mat#1{\ul{\ul{#1}}}
\def\vn{\vec{n}}

\def\expectCl#1{\ll #1\gg}

\usepackage{xcolor}

\DeclareGraphicsExtensions{.pdf,.eps}
\graphicspath{{figs/}}

\begin{document}

\title{Non-equilibrium nuclear spin distribution function in quantum dots subject to periodic pulses}

\author{Natalie J\"aschke}
\author{Andreas Fischer}

\address{Lehrstuhl f\"ur Theoretische Physik II, Technische Universit\"at Dortmund,
Otto-Hahn-Stra{\ss}e 4, 44227 Dortmund, Germany}

\author{E. Evers}
\author{V.V. Belykh}
\author{Alex Greilich}
\author{Manfred Bayer}
\address{Experimental Physik II, Technische Universit\"at Dortmund,
Otto-Hahn-Stra{\ss}e 4, 44227 Dortmund, Germany}

\author{Frithjof B.\ Anders}
\address{Lehrstuhl f\"ur Theoretische Physik II, Technische Universit\"at Dortmund,
Otto-Hahn-Stra{\ss}e 4, 44227 Dortmund, Germany}

\date{\today}

\begin{abstract}

Electron spin dephasing in a singly  charged semiconductor quantum dot 
can partially be suppressed by periodic laser pulsing. We propose 
a semi-classical approach describing the decoherence of the electron spin polarization 
governed by the hyperfine interaction with the nuclear spins as well as the probabilistic 
nature of the photon absorption. We use the steady-state Floquet condition to
analytically derive two subclasses of resonance conditions 
excellently predicting the peak locations in the part of the Overhauser field distribution which is projected in the direction of the external magnetic field. 
As a consequence of the periodic pulsing, a non-equilibrium distribution develops
as a function of time.
The numerical simulation of the coupled dynamics 
reveals the influence of the hyperfine coupling constant distribution 
onto the evolution  of the electron spin polarisation before the next laser pulse.
Experimental indications are provided for both subclasses of resonance conditions.

\end{abstract}
\maketitle

\section{Introduction}

Combining traditional  electronics with novel  spintronic  devices
has lead to an intensive investigation of semiconductor quantum dots
(QD) using electrical \cite{KastnerSET1992,Elzerman04}  or optical
probes \cite{PhysRevLett.96.227401,HansonSpinQdotsRMP2007}. High
localization of the electron wave function in the QD reduces the
decoherence facilitated by free electron motion,   but simultaneously
increases the hyperfine interaction strength between the confined
electron spin and the surrounding nuclear spins 
\cite{merkulov,CoishLossUnivers,FischerLoss2008,HansonSpinQdotsRMP2007}. 
Nevertheless, QD ensembles driven by periodic circular polarized
laser pump pulses provide a promising route for optically controlled
quantum functionality \cite{PhysRevLett.96.227401,Greilich28092007}.

A steady state of the spin system emerges from the periodic pulsing of
such QD ensembles that is substantially different to its equilibrium
starting point. Floquet's theorem provides a periodicity condition for this non-equilibrium steady state which translates into a mode-locking
resonance condition for the electron spin dynamics.  Early on, it was
conjectured that this mode-locking condition
\cite{PhysRevLett.96.227401} leads to a nuclei-induced frequency
focusing of electron spin coherence. Although the short time dephasing
remains unaltered, the resonance condition partially restores spin
coherence via constructive interference before the next laser pulse
arrives.

In this paper, we derive a semi-classical approach for the coupled
dynamics of electron spin and the nuclear spins in a single negatively
charged QD. The external magnetic field is applied in the Voigt
geometry, i.\,e.\ orthogonal to the optical axis, and the electron
spin is subject to periodic pulsing with a circular polarized
laser. The presented method takes into account the hyperfine
interactions between the electron and nuclear spins
\cite{merkulov,CoishLossUnivers,Glazov2012} as well as the Zeeman
terms for all spins. The method is used to calculate the emerging
non-equilibrium steady state induced by the periodic pulsing.

Our approach is based on the observation that the Overhauser field
generated by the large number of nuclei spin behaves as a classical
variable in leading order
\cite{merkulov,Glazov2012,stanekUhrig2013,Glasenapp2016}, particularly
in a large external magnetic field. Chen at al.\
\cite{PhysRevB.76.045312} showed that the quantum-dynamics of the
central spin model \cite{Gaudin1976,merkulov} can be accurately
approximated by expanding the path-integral representation around its
saddle-point, defined by a set of classical Euler-Lagrange equation of
motions. Subsequently, the quantum mechanical trace is replaced by a
configuration average over all classical spin configurations
\cite{PhysRevB.76.045312}.

The effect of laser pulses as well as the decay of the created trion,
however, requires a fully quantum mechanical treatment of the electron
spin dynamics reflecting the probabilistic nature of the photon
absorption and emission processes. In order to accommodate these
quantum effects, the unique correspondence between a
quantum-mechanical expectation value for a spin $1/2$ and the
components of the density matrix of such a spin subject to a classical
magnetic field is exploited.  The quantum nature of the spin-pumping
and the trion decay can therefore be included into an Ehrenfest
equation for the electron spin expectation value. It has almost the
same analytical structure as that of a classical spin, and the quantum
mechanics is encoded in the non-constant length of the classical spin
vector to account for the effect of each laser pulse and the
subsequent trion decay.

Although our theoretical simulations focus on a single QD, the
considerations can be extended to a QD ensemble. In a single QD, the
distribution of the hyperfine coupling constants is fixed while they
typically vary from QD to QD in a real ensemble.  Furthermore, the
different geometries of different QDs yield different electronic
confinement potentials and consequently, different laser energies are
required to pump the trion states of each QD \cite{Carter2009, Korenev2011}. A monotonous connection
between the trion excitation energy and the electron g-factor
$g_\mathrm{e}$ has been experimentally used
\cite{SpatzekGreilichBayer2011} to address sub-sets of a QD ensemble
with differently colored laser light.  Since the fluctuations of the
Overhauser field define the short-time dephasing time $T^*$
\cite{merkulov,HansonSpinQdotsRMP2007,PhysRevB.89.045317}, our
calculations can be interpreted as either simulations for an ensemble
of identical QDs, or the accumulated time average of many consecutive
measurements of the spin-polarisation on a single QD.  Experimentally,
the accumulated average of a QD ensemble is recorded
\cite{PhysRevLett.96.227401}. To account for this, we have to average
the results for a single QD over a typical distribution of
$g_\mathrm{e}$-factors as well as a distribution of times $T^*$ as
given in the experimental situation.

While the basic effect of the periodic pulsing was well understood in terms of a 
resonance condition for the electron spin dynamics  \cite{Greilich28092007},
a direct experimental access to 
the properties of the nuclear spin bath is absent under these conditions. 
One of the main objectives of this paper is to clarify the dynamics of 
the emerging mode-locking conditions
based on an analytical argument as well as a detailed analysis of the full
numerical simulation. We show that some of the basic features of
forming a non-equilibrium distribution function of the Overhauser field predicted by 
Petrov and Yakovlev \cite{petrovYakovlev}
prevails for the proper quantum mechanical treatment of the trion decay
based on a Lindblad approach.

Assuming a converged periodic Floquet state after infinitely many
laser pulses, we analytically derived two steady-state resonance
conditions: one is identical to the conjecture by Greilich et al.\
\cite{Greilich28092007} while the second condition additionally
depends on the ratio between the Larmor frequency and the trion decay
rate. It turns out that these analytic predictions excellently agree
with the full numerical simulations and can provide an upper bound of
the maximal achievable spin polarization in such experimental setups.

Recently, a fully quantum-mechanical treatment of the problem
\cite{BeugelingUhrigAnders2016} has addressed the question how the
nuclear non-equilibrium distribution function emerges due to the
periodic pump pulses of a QD. While Petrov and Yakovlev
\cite{petrovYakovlev} used simplified assumptions that each pump
pulse initializes the electron spin in a fully polarized state, the
quantum mechanical treatment of trion excitation and the subsequent
decay under reemitting a photon has been taken into account
\cite{BeugelingUhrigAnders2016}. A very slow growth of a peak
structure in an originally Gaussian Overhauser field distribution
\cite{merkulov,Glazov2012} has been reported where these emerging
peaks can be understood in terms of resonance conditions
\cite{BeugelingUhrigAnders2016,BeugelingUhrigAnders2017}.  In order to
address the dynamics of a reasonably large spin bath consisting of
$N=15-17$ nuclei, however, the hyperfine interaction was
perturbatively treated up to linear order in the spin-flip terms
during each pulse interval.  It remains unclear whether the slow
growth of the non-equilibrium distribution function is related to the
underestimation of the spin-flip term in the perturbation theory or is
already representative for a QD comprising typically $10^5$ nuclear
spins.

Since the characteristic frequency of the classical and the quantum
mechanical treatment of the spin precession are identical and given by
the effective Larmor frequency, we do not alter the relevant time
scale by resorting to a classical treatment of the individual nuclear
spins in order to treat (i) large numbers of nuclear spins and (ii)
allow for an isotropic dynamics induced by the hyperfine
interaction. We provide a simple scaling argument how to extrapolate
the time scales for the evolution of the non-equilibrium distribution
function to a realistic QD.

The theoretical simulations are augmented by experimental data recorded on an ensemble of $n$-doped quantum dots. Measurements address the magnetic field dependency of the mode locking amplitude as well
as the Fourier transform of the electron spin dynamics, both obtained by
Faraday rotation measurements
\cite{PhysRevLett.96.227401,Greilich28092007}. From the Fourier transforms with sufficient resolution we indeed find clear indications for precession modes fulfilling the predicted second class of resonance conditions that have not yet been observed before. These modes become particularly prominent around 4~T, where the mode-locked spin amplitude that can be assessed shortly before the impact of a pump pulse shows a minimum.


\subsection{Plan of the paper}
%
We use a semi-classical approximation to study the electron spin
dynamics and the development of a nuclear spin distribution in a
periodically pulsed QD system. A formalism for the simulation which
incorporates a classical description for the hyperfine interaction and
Larmor precession around the external magnetic field as well as for
the trion decay is presented. Sec.\,\ref{sec:theory} is divided in two
parts: one covers the theoretical basics, the other addresses the
methods used in the simulation. In the first part the central spin
model is introduced. In \ref{subsec:methods} the Lindblad formalism
for the trion decay is discussed. In the following, coupled equations
of motion are found for the electron spin and the nuclear spin bath. A
classical approach for the trion decay can be derived from the
Lindblad formalism. Under the assumption of a frozen Overhauser field
two sets of resonance conditions can be found. Lastly, the influence
of the Overhauser field on the electron spin is introduced.
Sec. \ref{sec:results} contains the results of the theoretical
simulations. We start with a brief introduction of the default
settings of the parameters. Those will be used to gain a fundamental
understanding of the time evolution of the system.  The rest of the
section is devoted to the variation of parameters like the external
magnetic field or the distribution of coupling
constants. Sec. \ref{sec:experimental.mode-spectrum} addresses the
results of experimental studies for the electron spin precession frequency spectrum, which show
clear indications for modes at both resonance conditions. The last
section will summarize the results and give an outlook to further
investigations.

\section{Models and methods}
\label{sec:theory}


We aim to describe a single electron charged QD subjected to periodic
laser pulses and to an externally applied magnetic field. The time
scales of the system vary greatly: time duration of the pulses \mbox{($\sim$
1.5\,ps)}, the trion decay \mbox{($\sim$ 0.4\,ns)} and the repetition
time of the pulse \mbox{(13.2\,ns)}\,\cite{Greilich28092007,
PhysRevLett.96.227401}. Therefore, the laser pumping will be treated
quantum-mechanically whereas we will use a semi-classical approach for
the trion decay and the electron spin dephasing between two
consecutive pulses.


\subsection{Central spin model}
\label{sec:CSM}

The Fermi contact hyperfine interaction between the central electronic
spin and the nuclear spins in the QD provides the largest contribution
to electron spin dephasing \cite{HansonSpinQdotsRMP2007} in a singly
charged semiconductor QD.  Other interactions such as dipole-dipole
interaction \cite{HansonSpinQdotsRMP2007} or the electrical
quadrupolar nuclear interactions are several orders of magnitude
smaller and, therefore, will be neglected in the following
\cite{merkulov,HansonSpinQdotsRMP2007,dyakonov}.

The Hamiltonian of the central spin model (CSM) accounts for the
effect of the external magnetic field on the electron and nuclear spins as
well as the hyperfine interaction between nuclear spin bath and
electron spin:
\begin{eqnarray}
\label{eqn:csm} 
H_\mathrm{CSM} &=&
g_\mathrm{e}\mu_\mathrm{B}\vec{B}_\mathrm{ext}\vec{\hat S} +
\mu_\mathrm{N} \vec{B}_\mathrm{ext}\sum_{k=1}^{N} g_k\vec{\hat I}_k 
\\
&& +
\sum_{k=1}^{N} A_k \vec{\hat I}_k \vec{\hat S} .  
\nonumber
\end{eqnarray}
%
The operators $\vec{\hat S}$ and $\vec{\hat I}_k$ denote
the electron spin and the $k^\mathrm{th}$ nuclear spin. $N$ labels the
number of nuclear spins.

All spins precess around the external magnetic field
$\vec{B}_\mathrm{ext}$ with the Larmor frequency
$\omega_\mathrm{e}=g_\mathrm{e}\mu_\mathrm{B}|\vec{B}_\mathrm{ext}|$
for the electron spin and
$\omega_\mathrm{N,k}=g_k\mu_\mathrm{N}|\vec{B}_\mathrm{ext}|$ for the
$k^\mathrm{th}$ nuclear spin, respectively. The last term in
\eqref{eqn:csm} encodes the hyperfine interaction between central spin
and nuclear spins via the Overhauser field, $\vec{\hat B}_\mathrm{N} =
\sum_k A_k \vec{\hat I}_k$. The feedback to the $k^\mathrm{th}$
nuclear spin is given by the Knight field $\vec{\hat B}_k = A_k
\vec{\hat S}$.  The strength of the coupling constants $A_k$ are
determined by the probability of an electron being present at the
position of the $k^\mathrm{th}$ nucleus $|\psi(\vec{R}_\mathrm{k})|^2$
\cite{merkulov,HansonSpinQdotsRMP2007}.

The fluctuations of the Overhauser field in absence of an external
magnetic field, $\expect{\vec{\hat B}_\mathrm{N}^2}$, define a
characteristic time scale of the system
\begin{align}\label{eqn:time_scale} (T^*)^{-2} = \sum_k A_k^2 \langle
\vec{\hat I}_k^2 \rangle
\end{align} governing the short-time electron spin decoherence.
Throughout the paper, we use $T^*$ as the characteristic time or
inverse energy scale in all calculations.  Note, that we have absorbed
$\hbar$ in the definition of time in the numerical calculations; At
the end, the time is converted back to physical units using
$T^*\approx 1$\,ns in order to connect with the experiments.

It is useful to introduce dimensionless coupling constants $a_k = T^*
A_k$ and dimensionless magnetic fields $|\vec{b}_\mathrm{ext}| =
g_\mathrm{e}\mu_\mathrm{B}T^* |\vec{B}_\mathrm{ext}|=
\omega_\mathrm{e}T^*$. This leads to the dimensionless Hamiltonian
\begin{align}
\label{eq:h-csm-nuclear-z} \bar H = H_\mathrm{CSM} T^* &=
\vec{b}_\mathrm{ext}\vec{\hat S} + z
\vec{b}_\mathrm{ext}\sum_k\vec{\hat I}_k + \sum_k a_k \vec{\hat I}_k
\vec{\hat S}
\end{align}
 where $z$ denotes the ratio between the nuclear Zeeman and
the electronic Zeeman energy. For simplicity, $g_k$ is taken as equal
for all nuclear spins.  For In$_x$Ga$_{1-x}$As QDs, $z =
\frac{g_k\mu_\mathrm{N}}{g_\mathrm{e}\mu_\mathrm{B}} \approx
(800)^{-1}$ replaces the small difference in the Ga,In and As Zeeman
energies by an averaged value $z$.  Recently, the effect of different
nuclear spin species on the dynamics has been investigated employing a
quantum mechanical perturbation theory
\cite{BeugelingUhrigAnders2017}. This requires a nuclei dependent
ratio $z_k$ in Eq.\ \eqref{eq:h-csm-nuclear-z} which is beyond the
scope of this paper.

In the experiments, the external magnetic field is applied in 
$x$-direction, in the Voigt geometry, while the laser beam direction,
which is perpendicular to this, defines the $z$-direction.

\subsection{Methods}
\label{subsec:methods}

The major challenge for the description of the pulse dynamics and the
build-up of a non-equilibrium steady state in a QD ensemble subject to
periodic laser pulses is the large separation of the time scales.
While the laser pulse duration typically is given by $T_\mathrm{P}=
1-4$\ ps, and can be treated as instantaneous to a good approximation,
the dephasing time due to the hyperfine interactions is three orders
of magnitude larger while the pulse repetition time is
$T_\mathrm{R}=13.2$\,ns in the experiments. Since the experiments are
performed at magnetic fields of the order of $1-6$T, the electronic
Larmor frequency $ |\vec{b}_\mathrm{ext}|$ is large compared to the
hyperfine interaction energy $1/T^*$.

Electronic spin polarization is generated by resonant circular
$\sigma^+$ laser pulses exciting the electron state $\ket{\uparrow}$
to a trion state $\ket{\uparrow\downarrow\Uparrow}$. Spin conservation and the formation of an electron singlett formation
prevent the excitation of the electron $\ket{\downarrow}$-state for
$\sigma^+$ circular polarisation.  The effective $g$ factor of the
trion is dominated by the hole spin and turns out to be negligibly
small. Therefore, precession of the trion state
$\ket{\uparrow\downarrow\Uparrow}$ to
$\ket{\uparrow\downarrow\Downarrow}$ in the external magnetic field is
omitted.  During the Larmor precession of the
$\ket{\downarrow}$-state, the trion state decays back to
$\ket{\uparrow}$ under emission of light at a decay time $1/\gamma$
which is typically $0.1-0.2T^*$. Clearly, this process must be treated
quantum mechanically by a Lindblad approach even though a simplified
approach has recently been proposed \cite{petrovYakovlev}.

The experimentally relevant time scales allow us to separate the time
evolution between two pulses into two steps: (i) the laser pulse which
is treated by an instantaneous unitary transformation of the
electronic part of the density operator; (ii) the decay of the trion
is accounted for by a Lindblad formalism and the simultaneous time
evolution of the coupled nuclear electronic system.

For the last step, one could remain within a fully quantum mechanical
description \cite{PhysRevB.89.045317,stanekUhrig2013} but is limited
to a relative small number of nuclear spins
\cite{PhysRevB.89.045317,BeugelingUhrigAnders2016}, or to short-time
dynamics \cite{stanekUhrig2013} using a TD-DMRG approach
\cite{Schollwoeck2011}.  Alternatively, one can map the dynamics onto
a set of classical equations of motion
\cite{merkulov,PhysRevB.76.045312,Fauseweh2017} which shows remarkably good
agreement with the full quantum mechanical treatment
\cite{stanekUhrig2013} but is easily extendable to a large number of
spins. Below, we address the key challenge of how to combine quantum
and classical calculations in a systematic way to incorporate the
formation on a non-equilibrium density distribution of the Overhauser
field \cite{petrovYakovlev} which is the origin of the self-focusing
experimentally observed by Greilich et al. \cite{Greilich28092007}.


\subsubsection{Lindblad approach}

We start from an Ising basis for the nuclear spins defined parallel to
the external field denoted by $\vec{m}=(m_1,...m_N)$ where $m_k$ is
the eigenvalue of $I_x$ for the $k^\mathrm{th}$ nuclear spin, and
$\sigma$ for the two spin orientations of the electron spin.  In that
basis, the matrix elements of the density operator of the coupled
nuclear-electronic system $\vec{\rho}(t)$ are denoted by
\begin{eqnarray} \rho_{(\sigma,\vec{m}),(\sigma',\vec{m'})}
(t)&=&\bra{\sigma,\vec{m}} \rho(t)\ket{\sigma',\vec{m}'}.
\end{eqnarray} 
Since the nuclear Zeeman energy as well as a single
coupling constant $a_k$ is very small, the nuclear spin configurations
can be treated as frozen on the very short time scale of the pulse
duration \cite{Greilich28092007}. For each frozen nuclear
configuration $\alpha=(\vec{m},\vec{m'})$, the basis of the electron
spin can be freely chosen. The states
$\ket{\uparrow},\ket{\downarrow}$ will denote the eigenvectors of
$\sigma_z$ with the eigenvalues $\pm 1$ while the electron spin Ising
basis parallel to the external field is assigned to
$\ket{\uparrow}_x,\ket{\downarrow}_x$.  Hence, we interpret
$\rho_{(\sigma,\vec{m}),(\sigma',\vec{m'})} (t)$ as matrix element of
mixed Ising bases: an Ising basis for the nuclear spins defined
parallel to the external field and an Ising basis for the electron
spins in the z-direction.

In this paper, we restrict ourselves to ideal $\pi$-pulses which
instantaneously excite a trion state
$\ket{\uparrow\downarrow\Uparrow}$ from an electron state
$\ket{\uparrow}$. Such an ideal pulse can be described by the unitary
transformation
\begin{eqnarray}
\label{eqn:unitary-pulse-operator} \hat{T}&=& \mathrm{i}
\ket{\uparrow\downarrow\Uparrow}\bra{\uparrow} +
\mathrm{i}\ket{\uparrow}\bra{\uparrow\downarrow\Uparrow} +
\ket{\downarrow}\bra{\downarrow}
\end{eqnarray} 
converting the initial density operator
$\rho^\mathrm{bp} $ to $\rho^\mathrm{ap} = \hat{T} \rho^\mathrm{bp}
\hat{T}^\dagger$.  Since the pulse only affects the electronic
subsystem, this transformation holds for each frozen nuclear
configuration $\alpha$ 
independent transformations
\begin{eqnarray}
\label{eq:pulse-II} \rho^\mathrm{ap}_\alpha &=& \hat{T}
\rho^\mathrm{bp}_\alpha \hat{T}^\dagger
\end{eqnarray} 
where $D$ is the dimension of the Hilbert space of the
nuclear spin bath.  In Eq.\ \eqref{eq:pulse-II},
$\rho^\mathrm{bp}_\alpha $ and $\rho^\mathrm{ap}_\alpha$ denote
$3\times 3$ matrices in the enlarged electronic Hilbert space
including the trion state $\ket{\uparrow\downarrow\Uparrow}$ -- for
details see appendix \ref{app:A}.

The trion decays under emission of a photon which is accounted for by
the Lindblad equation \cite{CarmichaelQuantumOpticsI}
\begin{align}
\label{eqn:linblad} \dot{\rho} = \mathcal{L}\rho(t) =
-\mathrm{i}[H_\mathrm{S}, \rho] - \gamma( s_2s_1 \rho + \rho s_2s_1 -
2s_1\rho s_2) .
\end{align} 
The second term describes the trion decay into the
electron state $\ket{\uparrow}$ by a constant decay rate $\gamma$
where the two transition operators, $s_1$ and $s_2$, are given by the
projectors $s_1 := \ket{\uparrow}\bra{\uparrow\downarrow\Uparrow}$ and
$s_2 := \ket{\uparrow\downarrow\Uparrow}\bra{\uparrow}$.  In an exact
treatment of the CSM, the system Hamiltonian $H_\mathrm{S}$ would be
$H_\mathrm{CSM}$. In the frozen Overhauser field approximation (FOA)
$H_\mathrm{S}$ only accounts for the electronic degrees of freedom.

Clearly, the Lindblad equation cannot be solved exactly for a CSM
comprising of large numbers of nuclear spins since the Hilbert space
grows exponentially. We are either restricted to small nuclear system
sizes \cite{PhysRevB.89.045317,BeugelingUhrigAnders2016} or we employ
the frozen nuclear approximation \cite{merkulov}, and arrive at
independent Lindblad equations
\begin{eqnarray}
\label{eqn:linblad-alpha}
\dot{\rho}_\alpha (t) = \mathcal{L}_\alpha \rho_\alpha(t) 
\end{eqnarray}
where the Liouvillian $\mathcal{L}_\alpha$ in each Overhauser field
configuration is defined by the system Hamiltonian
$H_\mathrm{S}(\alpha) =
g_\mathrm{e}\mu_\mathrm{B}\vec{S}\vec{B}_\mathrm{ext} +\Delta
H(\alpha)$ which describes the electronic precession in the external
magnetic field and a static, configuration dependent Overhauser
field. The trion decay in the Liouvillian is independent of the
nuclear bath configuration.  Eq.\ \eqref{eqn:linblad-alpha} can be
formally solved via
\begin{align}
\label{eq:lindblad-alpha} \rho_\alpha(t) =
\mathrm{e}^{\mathcal{L}_\alpha(t-t_0)}\rho_\alpha(t_0).
\end{align}

\subsubsection{Semi-classical approximation (SCA)}

The requirement to solve $D^2$ matrix equations
\eqref{eq:lindblad-alpha} in the frozen nuclear field approximation
drastically limits the number of bath spins which can be included in a
numerical simulation \cite{BeugelingUhrigAnders2016} to $N<20$. For
large numbers of nuclear spins contributing to an Overhauser field of
a finite length, however, the central limit theorem has been used to
calculate very accurately the short-time dynamics of the spin-spin
correlation function using a Gaussian distributed statical classical
Overhauser field \cite{merkulov}.

Chen et al.\ systematically derived corrections to the frozen
Overhauser field approximation \cite{PhysRevB.76.045312} starting from
the quantum mechanical path integral formulation of the problem.  The
path integral for expectation values uses spin coherent states for each
spin which are parameterized by the solid angle. The saddle point
approximation leads to $(N+1)$ coupled Euler-Lagrange equations
\cite{PhysRevB.76.045312,stanekUhrig2013,Fauseweh2017}
\begin{subequations}
\label{eq:SCA-EOM}
\begin{eqnarray}
\label{eq:central-spin} \dfrac{\mathrm{d}}{\mathrm{d}t} \vec{S} &=&
\left(\vec{b}_\mathrm{N} + \vec{b}_\mathrm{ext} \right)\times \vec{S}
\\
\label{eq:nuclear-spin-k} \dfrac{\mathrm{d}}{\mathrm{d}t} \vec{I}_k
&=& \left(a_k \vec{S} + z \vec{b}_\mathrm{ext}\right) \times
\vec{I}_k.
\end{eqnarray}
\end{subequations} 
with a remaining integral over all possible initial
spin configurations.  These equations describe the dynamics of coupled
classical spin vectors representing the central spin $\vec{S}$ and the
nuclear spin $\vec{I}_k$ by classical vectors.  Neglecting the
dynamics of the nuclear spins given by Eq.\ \eqref{eq:nuclear-spin-k}
recovers the FOA of Merkulov et al.\
\cite{merkulov,Glazov2012,Glazov2013:rev}, where the average over all
initial nuclear spin configurations has been replaced by a
configuration average over a Gaussian distributed Overhauser field
entering Eq.\ \eqref{eq:central-spin}.

A word is in order concerning the spin length. While the quantum
mechanical electron spin has $S=1/2$ and also a spin length of
$I=1/2$ is assumed for the nuclear spins we use a classical spin
vector of $|\vec{I}_k|=1$ in the numerical simulations below.
Clearly, Eq.\ \eqref{eq:central-spin} remains unaltered after
replacing $\vec{S}\to \vec{S'}=\vec{S}/S$.  In Eq.\
\eqref{eq:nuclear-spin-k}, we replace $\vec{I}_k\to \vec{I}_k/I$ to
justify the classical spin vector of $|\vec{I}_k|=1$. This
requires $a_k \vec{S}\to (S a_k) \vec{S}/S$ and $\vec{b}_\mathrm{N}= I
\sum_k a_k \vec{I}_k/I$. The modified equations of motion of the SCA
are then given by
\begin{subequations}
\label{eq:SCA-EOM-S-I-1}
\begin{eqnarray}
\label{eq:central-spin-one} \dfrac{\mathrm{d}}{\mathrm{d}t} \vec{S'}
&=& \left(\vec{b}_\mathrm{N} + \vec{b}_\mathrm{ext} \right)\times
\vec{S'} \\
\label{eq:nuclear-spin-k-one} \dfrac{\mathrm{d}}{\mathrm{d}t}
\vec{I'}_k &=& \left(a'_k \vec{S'} + z \vec{b}_\mathrm{ext}\right)
\times \vec{I'}_k.  \\
\label{eq:Bn-classic-one} \vec{b}_\mathrm{N} &=& I \sum_k a_k
\vec{I'}_k
\end{eqnarray}
\end{subequations} 
where all primed spin vectors are classical vectors
of length one.  As long as the Overhauser field $\vec{b}_\mathrm{N}$
remains unaltered, the electron spin Larmor frequency is invariant of
the spin length. Eq.\ \eqref{eq:Bn-classic-one} reveals that the
fluctuation of the Overhauser field is proportional to the spin
length, classically $I^2$ and quantum mechanically $I(I+1)$
\cite{merkulov} which becomes identical for large $I$. Defining
$\omega_{\rm fluc} = \sqrt{\ll\vec{B}_N^2 \gg}$ the spin lengths can
be absorbed into the definition of $T^*$ or $\omega_{\rm fluc}$
\cite{merkulov,hackmannPRL2015}. For $S = I$ the change of the
classical spin length leads to a modified coupling constant $a'_k =
a_k/2$ in Eq.\ \eqref{eq:nuclear-spin-k-one} and Eq.\
\eqref{eq:Bn-classic-one}.

The averaging over all Overhauser field configurations has been
interpreted as an averaging over an ensemble of identical QDs
\cite{PhysRevB.85.041303} each characterized by a classical spin
vector.  In experiments, however, variations of the QD size leads to
different characteristic time scales $T^*$.  Therefore, we view the
averaging procedure as being equivalent to averaging over repetitive
measurements \cite{Greilich28092007} and will perform the averaging
over different $T^*$ values of the QDs in the ensemble in a second
step.

\subsubsection{Combining the Lindblad and the semi-classical approach}

In order to connect the quantum mechanical treatment of the pulsed excitation and
the stochastic decay of the trion with a semi-classical description between the pulses,
we recall that the trace over the full Hilbert space was replaced by
an integration over all initial spins in the SCA. We discretize the
integration over all initial spins by generating $N_C$ configurations
comprising $N$ different nuclear spins and one central spin, each
equally distributed over the Bloch sphere, each weighted by a factor
$1/N_C$.

Since the thermal energy in the experiments typically exceeds all
other energy scales of $H_{\rm CSM}$, the initial quantum mechanical
density matrix $\rho_0$ is isotropic and proportional to the unity
matrix: $\rho_\alpha(t=0) = (1/2) \mat{1} (1/D)
\delta_{\vec{m},\vec{m'}}$.  By resorting to an average over $N_C$
classical configurations, we essentially replace the factor $(1/D)
\delta_{\vec{m},\vec{m'}}$ by the factor $1/N_C$ and identify the
label $\alpha$ by the classical configuration index,
\begin{eqnarray} 
\expect{\vec{\hat S}} &=& \frac{1}{D} \sum_{\alpha}
\Tr{\bar \rho_\alpha \vec{\hat S}} \nonumber \\ &\approx&
\frac{1}{N_C} \sum_{\mu} \Tr{ \rho_\mu \vec{\hat S}} \, = \, \,\,
\expectCl{\vec{S}}
\end{eqnarray} 
where the trace is calculated with the $2\times 2$
density matrix $\rho_\mu= \bar \rho_\alpha = D\times \rho_\alpha$
whose initial value is $(1/2)\mat{1}$. In the second line of the
equation, $\mu$ labels the classical configuration and
$\expectCl{\cdots}$ denotes the configuration average.

In a purely classical simulation, the classical spin $\vec{S}_\mu$ is
averaged directly.  Interpreting a classical spin vector $\vec{S}$
with $|\vec{S}| = 1/2$ as expectation value of a quantum mechanical
spin $1/2$ uniquely defines the corresponding $2\times 2$ density
matrix
\begin{eqnarray}
\label{eq:rho-spin} \rho_S &=& \left(
\begin{array}{cc} \frac{1}{2}+ S_z & S_x -i S_y \\ S_x + i S_y &
\frac{1}{2} - S_z
\end{array} \right).
\end{eqnarray} 
While a purely classical spin has a fixed length, the
quantum mechanical expectation value $\vec{S}_\mu$,
\begin{eqnarray} 
\vec{S}_\mu &=& \Tr{\rho_\mu \vec{\hat S}},
\end{eqnarray} 
can have arbitrary length reflecting the requirement
for a quantum ensemble description: the effect of the laser pump pulse
is an inherent statistical process.
 
Since classical equations of motion \eqref{eq:SCA-EOM} are norm
conserving for any vector $\vec{S}$, the restriction of a fixed spin
length of the central spin is not required. The classical equations of
motion \eqref{eq:SCA-EOM} only faithfully replace the unitary time
evolution of a quantum system under the influence of $H_{\rm CSM}$.

This unitary time evolution, however, is violated by the Lindblad
equation.  It accounts for the build-up of spin polarization due to
the laser pulse and consecutive trion decay: The length of the spin
expectation value quantum mechanically calculated with $\bar
\rho_\alpha$ will result in different spin polarizations from the
initial spin length. This reflects the fact that even an initially
pure quantum mechanical state typically will end up in a mixed state
after the trion decay.

The quantum mechanical evolution of the electronic density matrix
including the trion decay in a static magnetic field is determined by
Eq.\ \eqref{eqn:linblad}.  This requires the solution of eight
differential equations for the $3 \times 3$ matrix since the trace
remains conserved at all times.  We will show below, that these
equations are partially decoupled and are equivalent to those of the
spin and trion expectation values.

In order to connect the quantum mechanical treatment of the laser
pulse with the semi-classical equations of motion \eqref{eq:SCA-EOM},
we start from the FOA, i.\ e.\ treat $\vec{b}_\mathrm{N}$ as
static. After the laser pulse, the expectation value of any given
local observable $\hat O$ in the electronic subspace can be calculated
from the dynamics of the density matrix \eqref{eqn:linblad}:
\begin{eqnarray} 
\frac{d}{dt} \expect{\hat O} &=& \mathrm{i}
\Tr{\rho(t) [H_\mathrm{S}, \hat O] } -\gamma \Tr{\Delta\rho_L \hat O}
\label{eqn:general-O}
\end{eqnarray} where
\begin{eqnarray} 
\Delta\rho_L &=&
\ket{\uparrow\downarrow\Uparrow}\bra{\uparrow\downarrow\Uparrow}
\rho(t) +\rho(t)
\ket{\uparrow\downarrow\Uparrow}\bra{\uparrow\downarrow\Uparrow}
\nonumber \\ && -2\ket{\uparrow}\bra{\uparrow} P_{\rm T}(t)
\end{eqnarray} 
and $P_{\rm
T(t)}=\bra{\uparrow\downarrow\Uparrow}\rho(t) \ket
{\uparrow\downarrow\Uparrow}$ denotes the trion occupation
probability.  It is straight forward to derive the equation of motion
for the electron spin expectation values
\begin{eqnarray} 
\frac{d}{dt} \expect{\vec{\hat S}}_\mu &=& \vec{b}
\times \expect{\hat{ \vec{S}}}_\mu + \gamma P_{\rm T \mu}(t) \vec{e}_z
\label{eqn:spin-lindblad-expecation-value}
\end{eqnarray} 
which has a very intuitive interpretation: While the
trion decays back into the spin-up state contributing only to the spin
polarization in $z$-direction, the electronic spin precesses around
the effective magnetic field $\vec{b}=\vec{b}_\mathrm{N} +
\vec{b}_\mathrm{ext}$.

The solution of this set of equations requires the dynamics of the source term determined 
by the differential equation
\begin{eqnarray}
\frac{d}{dt} P_{\rm T,\mu }(t) &=& -2\gamma  P_{\rm T,\mu }(t)
\end{eqnarray}
that also is derived from \eqref{eqn:general-O}. It has the simple analytic solution
\begin{eqnarray}
P_{\rm T,\mu }(t) &=& P_{\rm T,\mu }(0) e^{-2\gamma t}
\end{eqnarray}
where $P_{\rm T,\mu }(0)$ is the trion occupation directly after the laser pump pulse.
These define the first four equations determining the evolution of the
nine matrix elements of the quantum mechanical density operator.

Since the trace is conserved, there are four more differential
equations required for the full solution of the density matrix. The
remaining four other differential equations only involve trion
off-diagonal matrix elements and also have a trivial exponential
decaying solution. Furthermore, these off-diagonal matrix elements do
not couple to the differential equations determining the spin dynamics
and can be neglected.
 
Consequently, we can include the Lindblad decay into the SCA replacing
\eqref{eq:central-spin} by
\begin{eqnarray} \dfrac{\mathrm{d}}{\mathrm{d}t} \vec{S}(t) &=&
\left(\vec{b}_\mathrm{N} + \vec{b}_\mathrm{ext} \right)\times
\vec{S}(t) + \gamma P_{\rm T \mu}(0) \vec{\mathrm{e}}_z e^{-2\gamma t}
\label{eq:full-central spin}
\end{eqnarray} 
where the time $t$ is measured relative to the last
pulse.

Within the FOA, this differential equation can be even solved analytically.
Without the source term, the homogeneous solution reads \cite{merkulov}
\begin{eqnarray}
\vec{S}_{\rm hom} &=& (\vec{A} \vn) \vn + [\vec{A}- (\vec{A}\vn) \vn] \cos(\w_L t)
\nonumber \\
&&
+\vn\times [\vec{A}- (\vec{A}\vn) \vn] \sin(\w_L t)
\end{eqnarray}
where the Larmor frequency is given by $\w_L=|\vec{b}| =|\vec{b}_\mathrm{N} +
\vec{b}_\mathrm{ext}|$, and $\vn = \vec{b}/\w_L$ denotes the unit vector in the direction
of the effective magnetic field. The solution is parametrized by the three component vector $\vec{A}$
which would be equal to $\vec{S}(0)$ in the absence of the source term.

The inhomogeneous solution has the form
\begin{eqnarray}
\vec{S}_{\rm in} &=& \vec{C} e^{-2\gamma t}.
\end{eqnarray}
Defining the rotation matrix $\mat{M}$ such that $\mat{M}\vec{v} = \vn\times \vec{v}$, we obtain
\begin{eqnarray}
 \vec{C}_\mu &=& -\frac{P_{\rm T \mu}(0)}{2} \left[  \mat{1} + \frac{\w_L}{2\gamma} \mat{M}_\mu
 \right]^{-1} \vec{e}_z.
\end{eqnarray}
From the total solution $\vec{S}_{\mu }(t) =\vec{S}_{\rm hom,\mu }(t) + \vec{S}_{\rm in,\mu }(t) $
and the initial condition we determine the
\begin{eqnarray}
\vec{A}_\mu  = \vec{S}_\mu(0) - \vec{C}_\mu 
\end{eqnarray}
where $\vec{S}_\mu(0)$ is the electronic spin expectation value of the
configuration $\mu$ after the pulse.  Since $\vec{C}_\mu$ has a negative sign, the spin 
polarization grows from $|\vec{S}_\mu(0)|$ directly after the pulse to $|\vec{A}_\mu|$, 
once the trion is completely decayed.

Within the SCA, we can even relax the constraint of a constant
Overhauser field by allowing $\vec{b}_\mathrm{N} \to
\vec{b}_\mathrm{N}(t)$. Then, the feedback of the central spin onto
the nuclear spins and visa versa is included at any time. But the
analytic solution derived above is no longer valid.

Let us summarize the individual steps of our hybrid quantum-classical
approach to a QD subject to periodic laser pulses.  Initially (i) we
generate $N_C$ classical spin configurations labeled by $\mu$ and
comprising a central spin and $N$ nuclear spins, each equally
distributed over the Bloch sphere.  (ii) We freeze the nuclear spins
and convert each central spin $\vec{S}_\mu$ of the classical
configuration into a $2\times 2$ density matrix $\rho_\mu$ using Eq.\
\eqref{eq:rho-spin}. This matrix is extended to a $3\times 3$ matrix
spanned by the enlarged Hilbert space including the trion.  (iii)
Then, the laser pulse is applied, described by Eq.\
\eqref{eq:pulse-II}, and quantum mechanical expectation values
$\vec{S}_\mu$ and $P_{T}$ are calculated directly after the pulse,
which define the initial conditions for solving the coupled equations
\eqref{eq:nuclear-spin-k-one}, \eqref{eq:Bn-classic-one} and
\eqref{eq:full-central spin} for the time interval up to
$t=T_\mathrm{R}$. For the next pulse, we go back to step (ii).  In
order to calculate expectation values, we average the quantity of
interest over all configurations $\mu$ for the given time $t$.

Our quantum-classical hybrid approach clearly reveals, that by the
necessary quantum mechanical treatment of the laser pulses the
simplified quantum to classical mapping of the spin degree of freedoms
does not hold for the electron spin. $\vec{S}$ loses its classical
interpretation even within a single configuration. The requirement for
a density matrix description has its deeper root in the statistical
nature of the photon absorption which is linked to the quantum
efficiency of the process. Although we only consider resonant photon
absorption, the theory can be simply extended to non-resonant
absorption by replacing $T$ in Eq.\ \eqref{eq:pulse-II} by the
appropriate unitary time evolution operator.

\subsubsection{Resonance conditions}
\label{sec:resonance-condition}

Before we present the full numerical solution in Sec.\
\ref{sec:results}, we analytically extract a steady-state solution
from the differential equation \eqref{eq:full-central spin} using
simplified approximations. As we demonstrate below, the resonance
conditions obtained in such a way agree remarkably well with our
simulations providing an a posteriori justification of these
simplifications.

Since the central spin dynamics is much faster than the nuclear spins,
we treat the nuclear spin dynamics as nearly frozen on the time scale
$T_\mathrm{R}$, i.\ e.\ the nuclear Zeeman term is neglected.
Furthermore, only the $x$-component of the Overhauser field is taken
into account since the $y,z$ components can be viewed as small
perturbations transversal to the large external magnetic field.

For the $\pi$ pulses discussed in this paper, we relate the electron
spin expectation values prior to the $N_\mathrm{P}$-th
pulse, $\vec{S}^\mathrm{bp}(N_\mathrm{P}T_R)$, to the one after the laser
pulse,
\begin{eqnarray}
\label{eqn:electron-spin-polarization}
\vec{S}^\mathrm{ap}(N_\mathrm{P}T_\mathrm{R}) &= &\left(0,0,
\frac{1}{2}\left(S^\mathrm{bp}_z-\frac{1}{2}\right)\right)^\mathrm{T}
\end{eqnarray} 
by applying the pulse operator
\eqref{eqn:unitary-pulse-operator}. The corresponding trion occupation
probability $P_{\mathrm{T}, \mu}(0) = (S^\mathrm{bp}_z + \frac{1}{2})$
is generated by the spin-up component of the density matrix.
Therefore, the trion and the electron spin state after each pulse
depend only on $S^\mathrm{bp}_z$, the $z$-component of the electron
spin right before the pulse.

These conditions are inserted into the analytical solution for the spin-expectation
values derived above:
\begin{align}
\label{eqn:centralSpinDyn}
\vec{S}(t) = \begin{pmatrix}
0 \\
-A_z \sin(\omega_\mathrm{L} t) + A_y \cos(\omega_\mathrm{L} t) - A_y \mathrm{e}^{-\overline{\gamma} t}\\
A_y \sin(\omega_\mathrm{L} t) + A_z \cos(\omega_\mathrm{L} t) - A_y \frac{\overline{\gamma}}{\omega_\mathrm{L}}\mathrm{e}^{-\overline{\gamma}t}
\end{pmatrix}
\end{align}
with $\overline{\gamma} = 2\gamma$ and the prefactors
\begin{align*}
A_y &= \dfrac{\omega_\mathrm{L} \overline{\gamma}}{\overline{\gamma}^2 + \omega_\mathrm{L}^2}\dfrac{2S^\mathrm{bp}_z + 1}{4}\\
A_z &= \dfrac{\overline{\gamma}^2}{\overline{\gamma}^2+\omega_\mathrm{L}^2}\dfrac{2S^\mathrm{bp}_z + 1}{4} + \dfrac{2S^\mathrm{bp}_z-1}{4}.
\end{align*}

In general, the static approximation of the Overhauser field is not
justified, since the effect of the Knight field on the nuclear spin is
required for the energy conservation law in the absence of the laser
pulses as well as the rearrangement of the Overhauser field
distribution as a function of time.  Since we are targeting the steady
state of the electron spin under periodic laser pumping, we (i) refer
to the Floquet periodicity condition for the $z$-component of the
electron spin
\begin{eqnarray}
\label{eqn:periodicity} S_z(T_\mathrm{R}) &=
&S^\mathrm{bp}_z
\end{eqnarray} 
and (ii) demand that the feedback of the Knight field
to the nuclear spins vanishes in average over the course of one pulse
repetition, i.\ e.\
\begin{align} 
\langle \dot{\vec{I}}_k\rangle_{T_\mathrm{R}} = \langle
a_k \vec{S} \times \vec{I}_k \rangle_{T_\mathrm{R}} = 0.
\end{align} 
For an almost static nuclear spin vector $\vec{I}_k$, this
translates into the vanishing of the average effect of the central
spin onto each nuclear spin over the pulse period $T_\mathrm{R}$,
\begin{align}
\label{eqn:avKnightField} \langle \vec{S}
\rangle_{T_\mathrm{R}} = \dfrac{1}{T_\mathrm{R}}
\int\limits_0^{T_\mathrm{R}} \vec{S}(t) \mathrm{d}t = 0
\end{align} 
independent of the coupling constant $a_k$. Note that the
electron spin lacks a $x$-component after the pulse, and this
component remains its zero value in a static effective magnetic field
in $x$-direction at all times.

Combining these two conditions with the analytic solution
\eqref{eqn:centralSpinDyn} reveals the $1/\omega_\mathrm{L}$
dependence of the averaged Knight field, see appendix \ref{app:B}, and
leads to the following equation
\begin{align}
\dfrac{\overline{\gamma}}{\omega_\mathrm{L}}\left(1-\cos(\omega_\mathrm{L} T_\mathrm{R})\right) - \sin(\omega_\mathrm{L} T_\mathrm{R}) = 0,
\end{align}
determining the set of Floquet values of the effective Larmor
frequency $\omega_\mathrm{L}$ under these assumptions.
Since the external magnetic field is fixed, the different values of
$\omega_\mathrm{L}$ translate to different steady-state values of the
Overhauser field in $x$-direction.

One class of solutions for $\omega_\mathrm{L}$ fulfils the resonance conditions 
\begin{align}
\label{eqn:even_rescond}
\omega_\mathrm{L} T_\mathrm{R} = 2\pi n \qquad \text{with}\quad n\in \mathbb{Z}
\end{align}
that was already discussed by Greilich\ et\ al.\
\cite{Greilich28092007}. They are only dependent on the external
magnetic field and independent of the trion decay rate. A second class
of solutions is determined by the transcendent equation
\begin{align}
\label{eqn:arctan_rescond}
\omega_\mathrm{L} T_\mathrm{R} = 2\arctan\left(\dfrac{\omega_\mathrm{L}}{\overline{\gamma}}\right) + 2\pi n \qquad n \in \mathbb{Z}
\end{align}
where the ratio of the Larmor frequency to the decay rate
$\overline{\gamma}$ generates an additional phase shift.  Since
$|\vec{b}_\mathrm{ext}| \gg |\vec{b}_\mathrm{N}|$ and the $\arctan$ is
monotonically increasing
$2\arctan\left(\frac{|\vec{b}_\mathrm{ext}|}{\overline{\gamma}}\right)$
serves as a good approximation.

For large external magnetic fields ($\omega_\mathrm{L} \gg
\overline{\gamma}$) the second class of solutions leads to Larmor
frequencies placed at odd resonance conditions, $\omega_\mathrm{L}
T_\mathrm{R} = \pi(2n+1)$, while for small magnetic fields these
additional peaks are brought closer to the even resonances.

In the central spin dynamics both Overhauser peak classes are
combined. The first class of solutions, defined by the even resonance
condition \eqref{eqn:even_rescond}, always is connected with an
electron spin that is aligned in negative $z$ direction right before
the pulse independent of the external magnetic field, $S^\mathrm{bp,
1}_z = -1$.  Then, the $\pi$ pulse has no effect on the electron spin
dynamics and $\vec{S}^\mathrm{bp}$ is identical to
$\vec{S}^\mathrm{ap}$.

Though the Larmor frequency $\omega_\mathrm{L}$ is strongly dependent
on the magnetic field for the second class of solutions, the spin
vector is always aligned in the positive $z$ direction, $S^\mathrm{bp,
2}_z = 1/3$.  The $\pi$ pulse leads to a flip of the CS from
$S^\mathrm{bp, 2}_z = 1/3$ to $S^\mathrm{ap, 2}_z = -1/3$.
Note that these are the only two polarizations $S^\mathrm{bp, 2}_z$
where the effect of the laser pulse conserves the spin length and
$|S^\mathrm{bp, 2}_z|$ is in a fixed point.

\subsubsection{Mode locked electron spin}
\label{sec:toymodel}

In order to set the stage for the analysis of the full numerical
simulations, we discuss the potential impact of the resonance
condition onto the central spin dynamics as well as the Overhauser
field distribution.  These Overhauser field distribution functions,
\begin{eqnarray}
p(b_{\mathrm{N},i}) &=&  \expectCl{\delta(B_{N,i}^{\mu}-b_{\mathrm{N},i})}
,
\end{eqnarray}
provide important statistical information about the nuclear spin
system, where the symbol $\expectCl{\cdots}$ denotes the configuration
average, and $i=x,y,z$.

Prior to applying the periodic laser pulses, we assume the system to
be in equilibrium and the high temperature limit to be valid, since
the thermal energy at $\sim 6$\,K is much larger than the hyperfine
interaction.  Therefore the nuclear spins can be regarded as
classical-spin vectors that are uniformly distributed on the unit
sphere. By using the law of large numbers this leads to Gaussian
distributed Overhauser fields $p(b_{\mathrm{N},i})$ in all spatial
directions \cite{merkulov}.

\begin{figure}[t]
\begin{center}
\includegraphics[width=0.45\textwidth,clip]{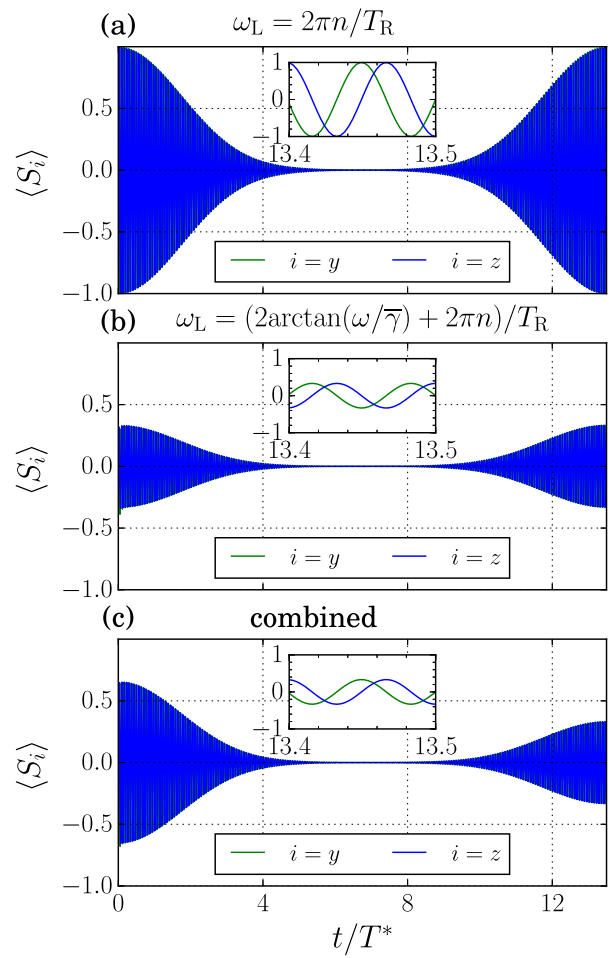} 
\end{center}
\caption{Toy model for the central spin dynamics. (a) and (b) give the central spin dynamics for one class of resonance conditions for $T_\mathrm{R} = 13.5\,T^*$. (c) combines the two classes with equal weight. The insets show the two oscillating electron spin components immediately before the next pulse.}
\label{fig:ToyModelRevival}

\end{figure}

To investigate the influence of the periodic pulse sequence on the
electron spin dynamics in a simplified toy model, we first combine the
precondition of the Gaussian envelope with the resonance condition
presented in the previous section. When the system reaches its steady
state, we assume that each class of resonance conditions leads to
$\delta$-peaks in $p(b_{\mathrm{N},x})$ inside a Gaussian
distribution.  Using Eq.\ \eqref{eqn:centralSpinDyn}, the solutions
for central spin dynamics for different Larmor frequencies are
superimposed and weighted according to the Gaussian envelope.  Within
the scope of this simple model we assume that both resonance
conditions contribute equally to the combined dynamics.

In order to relate the external field strength to the even resonance condition, we define $K'$ as
\begin{eqnarray}
\label{eqn:external-field-K}
K' &=&  \frac{|\vec{b}_\mathrm{ext}|  T_\mathrm{R}}{2\pi}.
\end{eqnarray}
For $K'\in \mathbb{Z}$, a free electron spin subject to an external
magnetic field $\vec{b}_\mathrm{ext}$ fulfills the resonance
condition. Off-resonance external magnetic fields can be quantified
via a deviation $\Delta K$ from the next integer value, i.\ e.\
$K'=K+\Delta K$ with $K\in \mathbb{Z}$.

In the example shown in Fig.\ \ref{fig:ToyModelRevival} we set $K=200$
which corresponds to a field strength of about $2$\,T. For such a
strong external field, the second class of resonance condition yields
peak positions at about $\pi(2n+1)/T_\mathrm{R}$. Note, that the
maximum length of the classical spin vector is 1.

The class of even resonance conditions leads to a central spin which
is aligned fully in the negative $z$ direction before the pulse. Hence
the electron spin polarization can fully be transferred to the next
pulse period since the $\pi$-pulse does not have any affect, and the
amplitude of the electron spin signal is maximal as shown in \mbox{Fig.\
\ref{fig:ToyModelRevival}(a).}

The electron spin configurations for the odd resonance conditions,
however, are aligned in positive $z$ direction. A full polarization of
the electron spin, however, is not possible according to Eq.\
\eqref{eqn:electron-spin-polarization}.
This sub-class also shows perfect revival as depicted in Fig.\
\ref{fig:ToyModelRevival}(b).
The perfect revivals in each sub-class at the end of the pulse period
being a consequence of the resonance condition, is destroyed by the
superposition of both since the spins point in opposite directions at
the end of the period. When weighting both sub-classes equally the
revival is significantly reduced as demonstrated in Fig.\
\ref{fig:ToyModelRevival}(c). The revival can be completely suppressed
when weighting the first and the second subclass in the ratio 1:3 --
not shown here.

\section{Results}
\label{sec:results}

\subsection{Distributions of hyperfine couplings} 
\label{couplingConstants}

While the short-time dynamics of the QD is governed by $T^*$ and
therefore independent of a particular $a_k$ distribution the long-time
dynamics is influenced by the probability density function $p(a_k)$ of
the coupling constants.  Several different distributions have 
been used for the CSM \cite{CoishLossUnivers,FischerLoss2008,gradedBoxModel2009,
stanekUhrig2013,PhysRevB.89.045317}, ranging from the simple box model
\cite{petrovYakovlev} which assumes equal coupling constants $a_k = a
= 1/\sqrt{N},\quad \forall k,$ to the more elaborate distributions of
coupling constants $p(a_k)$
\cite{HansonSpinQdotsRMP2007,CoishLossUnivers,FischerLoss2008,PhysRevB.89.045317}.

A simplifed constant distribution has advantages concerning
computation time whereas others provide a more realistic description
of the hyperfine coupling. The coupling constants are proportional to
the electronic probability of presence at the $k^\mathrm{th}$ nuclear
spin given by the envelope of the electron wave function
$\psi(\vec{R}_k)$,
\begin{align}
\psi(\vec{R}_k) \propto \exp\left(-\dfrac{1}{2}\left(\dfrac{r}{L_0}\right)^m\right),
\end{align}
at the location of the nucleus $\vec{R}_k$. $L_0$ is the
characteristic length scale of the QD and of the order of
$L_0\approx5$\,nm. For a spherical QD, a probability density function
\begin{align}\label{eqn:hackmanDist}
p(a) = -\dfrac{3}{m r_0^3}\dfrac{1}{a} \left(\ln\left(\dfrac{a_\mathrm{max}}{a}\right)\right)^{\frac{3-m}{m}}
\end{align}
has been derived \cite{PhysRevB.89.045317} where $r_0$ is the ratio
between an artificial cut-off $R$ and $L_0$. $a_\mathrm{max}$ is the
largest occurring coupling constant and contains information about the
underlying material. 
For $m=2$ the coupling constants are defined by $a = a_\mathrm{max}\exp(-r_0^2 x^{2/3})$.

For $m=3$ this distribution is related to exponential coupling constants $a =a_\mathrm{max}\exp(-r_0^3 x)$ with $x\in \mathcal{U}([0,1])$. These coupling constants, c.\,f. \cite{Seifert2016}, can also be calculated by 
\begin{align}
\label{eqn:expAk}
a_k = C e^{ -(k-1) \lambda},
\end{align}
with $k=1...N$ and \mbox{$C=\sqrt{\frac{1-\exp(-2\lambda)}{1-\exp(-2\lambda N))}}$}. $\lambda$ determines the spread of the coupling constants depending on the proportion of the volume of the quantum dot and the number of nuclear spins taken into account, $\lambda \sim r_0^3/N$.

\subsection{Definitions of the parameters}

The dynamics of the electron spin $\langle S_z \rangle$ and the
distribution of the Overhauser field $p(b_{\mathrm{N},i})$ with
$i=x,y,z$ in a system subjected to periodic laser pulses is
investigated. The parameters are chosen to correspond to the
experimental setup \cite{Greilich28092007}.

Unless stated otherwise, these parameters will stay the same in the
following sections where only one parameter is varied. We use a bath
size of $N=100$ nuclear spins and average over $N_\mathrm{C} = 10^5$
configurations. The length of the classical nuclear spin vector is
$|\vec{I}_k| = 1$. This is also the maximal length for the electron
spin vector $|\vec{S}_\mathrm{max}|=1$. Therefore, Eq.\
\eqref{eq:rho-spin} and Eq.\ \eqref{eq:full-central spin} have been
adjusted accordingly as discussed above.

For the theoretical simulations, we set the separation time between
two instantaneous pulses $T_\mathrm{R}= 13.5\,T^*$ for convenience
while the experimental constrains lead to $T_\mathrm{R}= 13.2\,$ns
\cite{Greilich28092007}. The trion decay rate is given by $\gamma =
10\,\frac{1}{T^*}$. We have used the conversion factor $T^*\approx
1$\,ns for simplicity to make contact with the experiments.  The scope
of this paper is to provide a basic understanding of the dynamics
observed in periodically driven QDs and not the fitting of a specific
experiment.

We convert $|\vec{b}_\mathrm{ext}|$ in a dimensionless number $K'$
defined in Eq.\ \eqref{eqn:external-field-K} to clearly signal a
resonance condition of the external magnetic field.
$b_\mathrm{ext}(K=200) \approx 93\,(T^*)^{-1}$ is equivalent to
$B_\mathrm{ext}\approx 2$\,T using the proper conversion constants.
The modification of $K'$ from an integer value to an arbitrary real
number $(K+\Delta K)$ can be used to understand deviations from the
resonance conditions which can also arise in a QD ensemble due to
different $g$ factors of individual QDs.

The strength of the nuclear Zeeman coupling is defined by the factor
$z = \frac{g_k \mu_\mathrm{N}}{g_\mathrm{e}\mu_\mathrm{B}} $ between
the nuclear and electron Zeeman energy as introduced in Eq.\
\eqref{eq:h-csm-nuclear-z}.  The coupling of the nuclear spins to the
external magnetic field can be explicitly neglected in the theoretical
simulations by setting $z=0$.

We begin with the so-called box model
\cite{BoxModel2009,petrovYakovlev}, i.\ e.\ we set all
$a_k=a=1/\sqrt{N}$ to reveal the basic properties of the dynamics
before presenting data obtained by numerically very expensive
simulations.  For nuclear spins coupling with individual $a_k$ to the
central spin, $N+1$ coupled equations \eqref{eq:SCA-EOM} have to be
solved. By using the box model the equations for the nuclear spins
collapse to a single EOM for the Overhauser field and the set of
equations is reduced to two.  We use the Runge-Kutta fourth-order
method to solve the differential equations. The step width has to be
adapted according to the strength of the external magnetic field to
resolve the Larmor precession of the central spin. For an external
field of $K=200$ a step width of $\sim 0.001 T^*$ has proven to be
sufficient.

We start with a completely unpolarized system. At $t=0$ the first
pulse is applied. The distribution of the Overhauser field is measured
immediately before the next pulse.

\subsection{Benchmarking the semiclassical equation of motion}

\begin{figure}[t]
\begin{center}
\includegraphics[width=0.45\textwidth,clip]{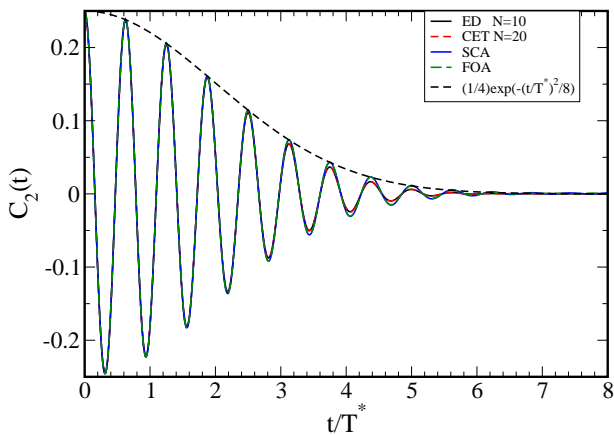} 
\caption{Comparison between spin correlation function $C_2(t)=\expect{S_z(t)S_z}$ in a finite
magnetic field $b_x=10$ calculated using different quantum mechanical approaches, the  
CET  with $N=20$ and ED  with $N=10$,
and the two classical approaches, SCA and FOA, with $N=100$ spins and $N_C=100,000$. The CET data have been taken from Fig.\ 7 in Ref.\ \cite{PhysRevB.89.045317}.
}
\label{fig:benchmark-C2}
\end{center}
\end{figure}

In order to benchmark the  quality of the SCA
\cite{Glazov2012,Glazov2013:rev,stanekUhrig2013} employed in this
paper, we compare the spin correlation function
$C_2(t)=\expect{S_z(t)S_z}$ in a finite magnetic field $b_x=10$
obtained with the two classical approaches, the SCA and FOA, with the
quantum mechanical results calculated using a Chebyshev expansion
technique (CET) \cite{Fehske-RMP2006} and via exact diagonalization
(ED) of the Hamiltonian.
Our results are summarized in Fig.\ \ref{fig:benchmark-C2}.  

Note that the dephasing time scale is governed by the fluctuation of
the Overhauser field $\langle\vec{B}_N^2\rangle$ \cite{merkulov} where
the average spin length enters. Since the quantum mechanical
simulations have been performed with $I=1/2$, we have absorbed the
difference in the Overhauser fluctuations between $\langle
\vec{I}_k^2\rangle = |\vec{I}_k|^2 = 1$ of the classical simulation
and $\langle \vec{\hat{I}}_k^2\rangle = I(I+1)= 3/4$ of the quantum
simulations into rescaled coupling constants $a_k\to \tilde a_k =
a_k\sqrt{3/4}$ for the SCA and the FOA.

While the SCA and the FOA provide identical results they also agree
remarkably well with the two quantum mechanical data sets for N=10
(ED) and $N=20$ (CET). The difference between classical and the
quantum simulations can be attributed to $1/N$ effects which are
suppressed using a large number of bath spins in a time-dependent
density matrix renormalization group approach \cite{stanekUhrig2013}.

Since the spin length is conserved in the SCA and the feedback
involves always a coupling constant, the differences can be absorbed
into the definitions of the coupling constants or the reference time
and energy scales, respectively. After establishing this quality of
the SCA, we use the energy and time scales as defined in Sec.\
\ref{sec:CSM} throughout the paper.  For a classical spin length of 1,
the coupling constants used are
\begin{eqnarray} 
a_k' &=& \frac{a_k}{2} =
\frac{A_k/2}{\sqrt{\sum_{k=1}^N \frac{1}{4} A_k^2}} =
\frac{A_k}{\sqrt{\sum_{k=1}^N A_k^2}} \,\, .
\end{eqnarray}

\subsection{Non-equilibrium Overhauser field distribution function: nuclear self-focusing}

\subsubsection{Influence of the number of pulses}
\label{subsec:numberPulses}

\begin{figure}[t]
\begin{center}
\includegraphics[width=0.4\textwidth,clip]{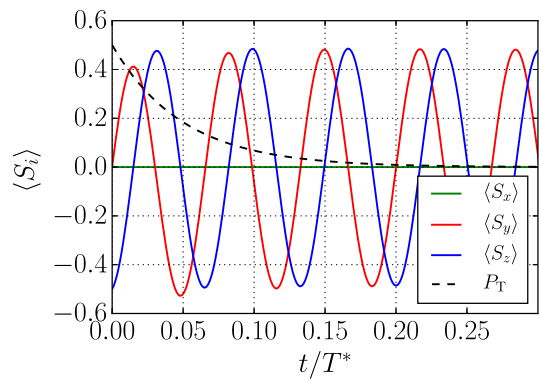} 
\caption{Precession of the electron spin after the first pulse. After the pulse the $x$ and $y$ component are zero and the $z$ component is given by $(S^\mathrm{bp}_z-1)/2$. 
The analytical solution of the exponential trion decay has been added as dashed line.
}
\label{fig:CSt_number_of_pulses_detail}
\end{center}
\end{figure}

The electron spin dynamics is dominated by the precession around the
strong external magnetic field. The electron spin component parallel
to the external magnetic field remains at approximately zero since the
laser pumping only generates a spin polarization in the $z$-direction.
The components perpendicular to the external magnetic field show the
electron spin precession as demonstrated in \mbox{Fig.\
\ref{fig:CSt_number_of_pulses_detail}}. The first pulse at $t=0$
depletes the $\ket{\!\uparrow}$ state of the previously unpolarized
electron spin. Therefore the electron spin starts precessing from
$\vec{S}(t=0) = \vec{S}^\mathrm{ap} = -0.5\ \vec{\mathrm{e}}_z$.  The
trion decay leads to a steady increase in the electron spin
polarization on a time scale of $0.1-0.2 T^*$.

\begin{figure}[tp]
\begin{center}
\includegraphics[width=0.4\textwidth,clip]{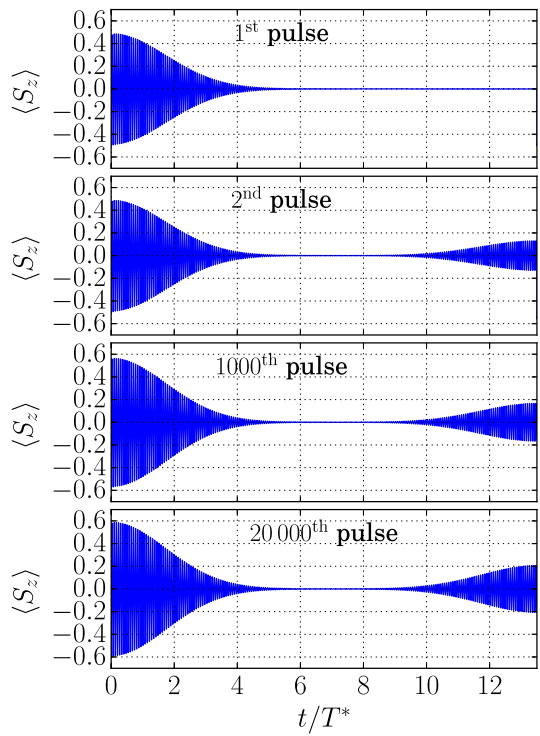}
\end{center}
\caption{Dynamics of the $z$ component of the electron spin during the
time interval between two pulses for four different numbers of pulses
calculated for the box model.}

\label{fig:CSt_number_of_pulses}

\end{figure}

While coherent oscillations are observed on a very short time scale,
defined by the inverse Larmor-frequency, the hyperfine interaction
leads to dephasing which is governed by $T^*$, see Fig.\
\ref{fig:CSt_number_of_pulses}. While the electron spin dephases
completely after the first pulse, we observe a revival of electron
spin polarization after the second pulse to an amplitude of $|\vec{S}|
\approx 0.14$ just before the next laser pulse arrives.  After that,
the central spin revival amplitude slowly grows with an increasing
number of pulses.

\begin{figure}[t]
\begin{center}
\includegraphics[width=0.47\textwidth,clip]{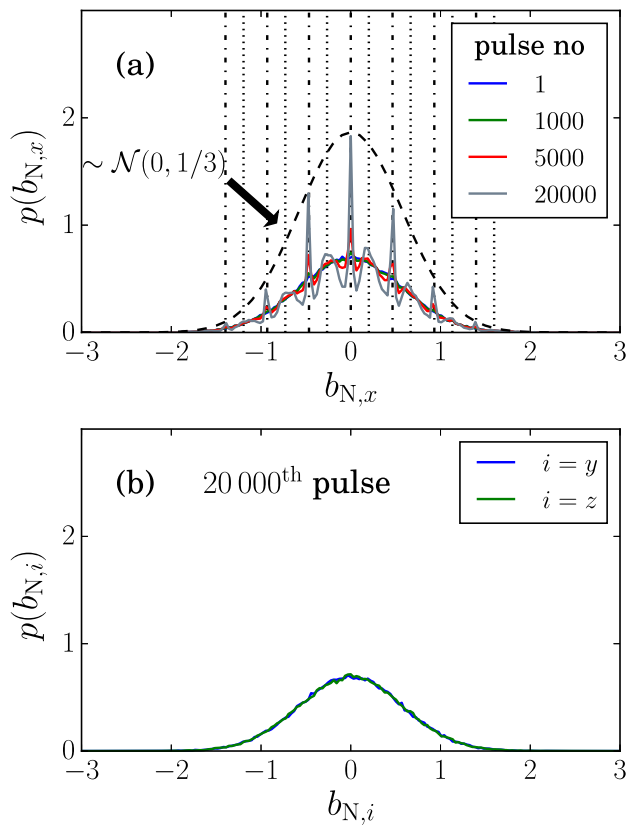}
\end{center}
\caption{Panel (a) shows the influence of the number of pump pulses on
the $x$ component of the density distribution of the Overhauser field
$b_{\mathrm{N},x}$ for $K=200$. At the beginning of the pulse sequence
$b_{\mathrm{N},x}$ is normally distributed. $b_{\mathrm{N},x} \sim
\mathcal{N}(0, 1/3)$. This scaled normal distribution is the envelope
of the emerging density distribution.  The vertical lines indicate the
two sub-classes of theoretical peak positions defined by the equations
\eqref{eqn:even_rescond} and \eqref{eqn:arctan_rescond}. (b) the
corresponding $y$ and $z$ component of $p(\vec{b}_\mathrm{N})$ after
20\,000 pump pulses. }
\label{fig:hist_number_of_pulses}
\end{figure}

The central spin dynamics is directly connected to the three
distributions of the Overhauser field $p(b_{\mathrm{N},x}),
p(b_{\mathrm{N},y}), p(b_{\mathrm{N},z})$.  The evolution of
$p(b_{\mathrm{N},x})$ with the number of pulses is shown for the box
model in Fig.\ \ref{fig:hist_number_of_pulses}. At $t=0$ the
Overhauser field is unpolarized, implying that all Overhauser field
components follow a normal distribution $\mathcal{N}(0,
(\langle\vec{I}_k^2\rangle/3 = 1/3))$.

If the system is subjected to periodic pump pulses the distributions
of the Overhauser field components perpendicular to the external
magnetic field do not change from the initial Gaussian
distribution. However, a new distribution emerges for
$b_{\mathrm{N},x}$.  Though the envelope of the distribution stays
Gaussian, peaks begin to emerge at pronounced positions that become
more distinct with time.  We have identified two sub-sets of
peaks. The distance between every other peak is given by the resonance
condition, $\Delta b_{\mathrm{N},x} = 2\pi/T_\mathrm{R}$.

Despite the strong approximations made in Sec.\
\ref{sec:resonance-condition} on the resonance condition, the peak
structure calculated in the fully numerical simulation of the EOM of
the SCA, shown in Fig.\,\ref{fig:hist_number_of_pulses}, agrees
remarkably well with the theoretical predictions for the resonance
condition which have been added as vertical dotted and dashed-dotted
lines in the figure. We only observe deviations of $1-2\,\%$ and up to
$9\,\%$ at most.

\subsubsection{Influence of the external magnetic field strength}
\label{Bext}

The external magnetic field has two functions: (i) it induces a
coherent oscillation of the spin polarization and (ii) it can also
suppress dephasing stemming from the long-time fluctuations of the
Overhauser field.  It has been shown, that the accuracy of the FOA
approximation \cite{merkulov} increases with increasing magnetic field
\cite{Glazov2012,PhysRevB.89.045317,Glasenapp2016}. Only in the theory
of higher order correlation functions additional processes have to be
included in order to make connection to the experiment
\cite{FroehlingAnders2017}.

\begin{figure}[t]
\begin{center}
\includegraphics[width=0.47\textwidth,clip]{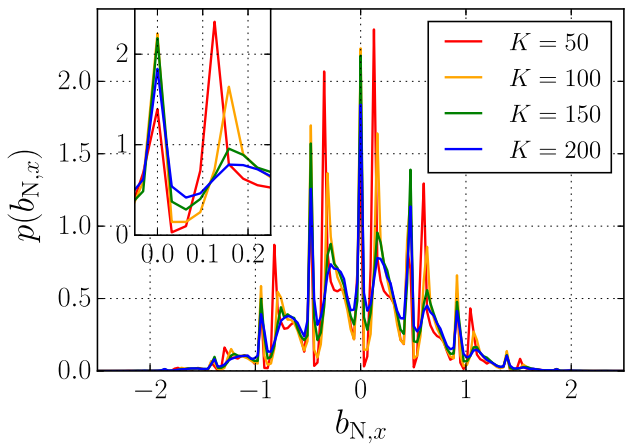}
\end{center}

\caption{Overhauser field distribution along the external magnetic
field for a different strength of $\vec{b}_\mathrm{ext} =
\frac{2\pi}{T_\mathrm{R}} K \vec{\mathrm{e}}_x$.  The inset shows the
peak for the even resonance condition at $b_{\mathrm{N},x}=0$ and the
shifted peak given by the ratio $\omega_\mathrm{L}/\overline{\gamma}$,
see Eq.\ \eqref{eqn:arctan_rescond}, after $20\,000$ pulses. }
\label{fig:hist_Bext}

\end{figure}

The strength of the external magnetic field plays an important role in
the development of the peak structure of the Overhauser field
distribution. In this subsection we examine the dependence on magnetic
fields as well as the resonance conditions Eq.\
\eqref{eqn:arctan_rescond} and Eq.\ \eqref{eqn:even_rescond}. A low
magnetic field allows for a fast build-up of the Overhauser field
distribution due to the $1/\omega_\mathrm{L}$ dependency of the Knight
field after integrating Eq.\ \eqref{eq:nuclear-spin-k}.

The Overhauser field distribution $p(b_{\mathrm{N},x})$ is plotted
for four different resonant magnetic field values $K=50$, $100$,
$150$ and $200$ after $20\,000$ pulses in Fig.\
\ref{fig:hist_Bext}. The inset focusses on one peak at $b_{N,x}=0$
belonging to the even resonance condition and one peak corresponding
to Eq.\ \eqref{eqn:arctan_rescond}.  While the even resonance peaks
located at positions independent on the external magnetic field
values, the peaks following \eqref{eqn:arctan_rescond} are shifted
away with increasing field strength $K$, as predicted by Eq.\
\eqref{eqn:arctan_rescond}.

The peak positions of the two classes of peaks are well described by
the analytical predictions. The additional features that become
apparent in the simulations cannot be derived from the analytical
results: the weight of each class of peaks in the combined Overhauser
field distribution. For strong external magnetic fields the peaks at
even resonance are still sharp, while the second sub-class of peaks
have a less distinct shape.

\subsubsection{Electron spin revival}

\begin{figure}[t]
\begin{center}
\includegraphics[width=0.5\textwidth,clip]{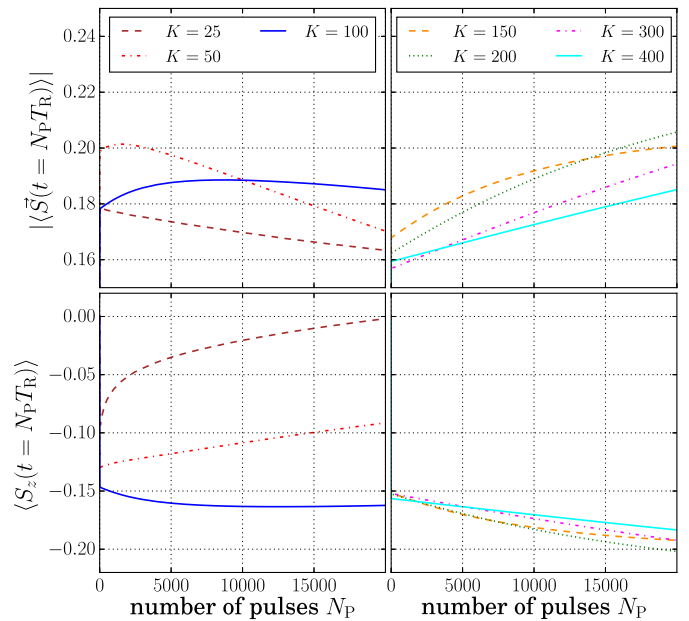}

\caption{Revival amplitude and $S_z$ component of the electron spin
for external magnetic fields $|\vec{b}_\mathrm{ext}(K)| = 2\pi
K/T_\mathrm{R}$ applied in the $x$ direction.  }
\label{fig:revival_Bext}
\end{center}
\end{figure}

\begin{figure}[t]
\begin{center}
\includegraphics[width=0.35\textwidth,clip]{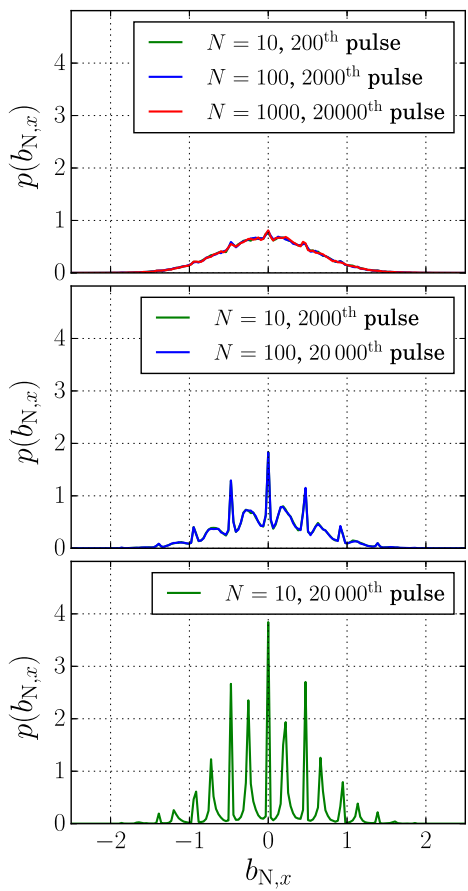}

\caption{Time evolution of the Overhauser field distributions parallel
to the external magnetic field for different numbers of nuclear
spins. The three panels show $p(b_{N,x})$ for three different but
fixed ratios $r=N_\mathrm{P}/N =20,200,2000$.  Parameters: $K=200$.  }
\label{fig:hist_NumberNuclearSpins}
\end{center}
\end{figure}

As an experimentally accessible quantity through the mode-locking amplitude, the electron spin revival
merits a more in-depth investigation. The electron spin dynamics is
intertwined with the Overhauser field distribution. It determines the
revival behavior since the superposition of configurations with
different Larmor frequencies suppresses the growth of the central spin
revival. That raises the question which properties of
$p(b_{\mathrm{N},x})$ influence the final revival amplitude.

The first class of peaks in the Overhauser field distribution is
independent of the external magnetic field for all integer values of
$K$. Since the period length of all even frequencies contributing to
the electron-spin precession fit into $T_\mathrm{R}$ as integer, the
central spin configurations are always aligned in negative $z$
direction before the pulse.

For configurations characterized by the second resonance condition,
the orientation of the electron spin prior to the next laser pulse
depends on the external magnetic field strength: Contrary to the
results of the simple toy model it can acquire a spin-polarization in
$y$-direction just before the next laser pulse that does not influence
the value of the spin-polarization after the pulse.  The contribution
of those configurations to the total signal is determined by the
spectral weight of the peaks in the distribution function that cannot
be obtained from the resonance condition.

Figure\,\ref{fig:revival_Bext} shows the influence of the external
magnetic field on the amplitude and the $z$ component of the electron
spin revival measured directly before the next pulse as function of
the pulse number $N_\mathrm{P}$.

For external magnetic field strengths $K\le 100$, the revival
amplitude decreases after the initial increase after the second
pulse. Here the phase shift leads to a alignment of the central spin
in the $y$ direction before the pulse which does not influence the
pumping process as is seen in Eq.\
\eqref{eqn:electron-spin-polarization}.

For larger magnetic fields, i.\ e.\ $K> 100$, 
the electron spin polarization is
aligned in $z$-direction.  Due to the mismatch in the probability
weight of the resonance conditions the revival increases.  The peaked
non-equilibrium Overhauser field distribution, however, emerges slower
for increasing magnetic fields due to the $1/\omega_L$ dependency of the
averaged Knight field, see Eq.\,\eqref{eqn:avKnightField}, leading to
a slower increase of the revival.  The electron spin polarization is
not yet converged after $20\,000$ pulses as seen in the two left panels in
Fig.\ \ref{fig:revival_Bext}.

\subsection{Scaling behavior of the number of nuclear spins}\label{scalingArg}

\begin{figure}[t]
\begin{center}

\includegraphics[width=0.45\textwidth,clip]{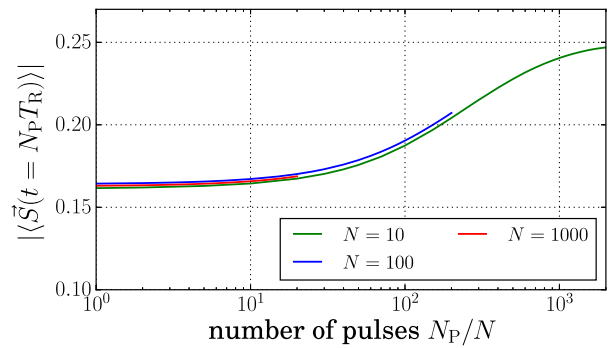}

\caption{Revival amplitude of the electron spin for different numbers
of nuclear spins.  Parameters as in Fig.\
\ref{fig:hist_NumberNuclearSpins}.  }
\label{fig:revival_NumberNuclearSpins}
\end{center}
\end{figure}

In the experimental setup the data are measured after an initial
pulsing period which lasts from a few seconds up to 20 min
\cite{Greilich28092007}.  For a laser repetition of $\sim 13.5$\,ns this
corresponds to $74 \times 10^6$ pulses per second.  Such large time
scales are impossible to achieve with our simulations even for the
simplified box model.  Therefore, it is useful to derive and exploit a
scaling relation associated with the number of nuclear spins in order to
extrapolate the possible steady-state of the system.

In Figure\ \ref{fig:hist_NumberNuclearSpins}, the time evolution of the
Overhauser field distribution for different numbers of nuclear spins
is shown: the larger the number of nuclear spins, the slower the
build-up of $p(b_{N,x})$.  The distribution $p(b_{N,x})$ is plotted
for different combination of $N$ and pulse numbers $N_\mathrm{P}$ for
a constant ratio $r=N_\mathrm{P}/N =20,200,2000$: $p(b_{N,x})$ is
universal and only depends on the ratio $r$.

This observed scaling behavior is attributed to the dependence of the
Knight field on the strength of the coupling constants, Eq.\
\eqref{eq:nuclear-spin-k}, and to the influence of the Overhauser
field on the central spin, Eq.\ \eqref{eq:central-spin}.  Since the
electron spin dynamics is fed back to the nuclear spins via the
coupling constant $a_k = 1/\sqrt{N}$ the build-up scales with $a_k^2
\propto 1/N$. Consequently, the slower feedback of the electron-spin
dynamics onto the $p(b_{N,x})$ with increasing number of nuclei must
be compensated by an additional number of laser pulses.

Although we have only demonstrated this scaling property for the box
model, we will show below that qualitatively similar scaling behavior
prevails for an arbitrary distribution function $p(a)$ when $T^*$ is
used as a reference time scale independent of $N$.  We will exploit
this scaling law to perform simulations with as little nuclear spins
as possible and extrapolate our results to the realistic number of
nuclear spins in a QD. The results obtained for $N=10$ nuclei and
20\,000 pulses are therefore equivalent to those of $10^5$ nuclei and
$2\cdot 10^9$ pulses, corresponding to approximately 2\,sec in a
typical experimental setup.

The amplitude of the electron spin revival for different numbers of
nuclear spins is depicted in Fig.\
\ref{fig:revival_NumberNuclearSpins} vs $r=N_P/N$ following the same
scaling law. Since the steady-state is approached but has not been
reached even for $N=10$ and 20\,000 pulses, we conjecture that we
would need another factor 10-100 more pulses to achieve final
convergence. This would translate to reaching the steady-state after
approximately half a minute to several minutes of pulsing which is in
the same order of magnitude as in the experiments
\cite{Greilich28092007}.

\subsection{Influence of an external magnetic field off resonance}
\label{offRes}

For a given applied external magnetic field and a fixed laser
repetition time $T_\mathrm{R}$, an individual QD may not fulfil the
resonance condition due to its electron $g_\mathrm{e}$ factor leading
to a non-integer value of $K'$ in \eqref{eqn:external-field-K}.  We
have introduced the parameter $\Delta K$ to represent the distance of
$K'$ to the closest integer value $K$ in order to measure the distance
from the integer resonance condition.

\begin{figure}[t]
\begin{center}
 \includegraphics[width=0.35\textwidth,clip]{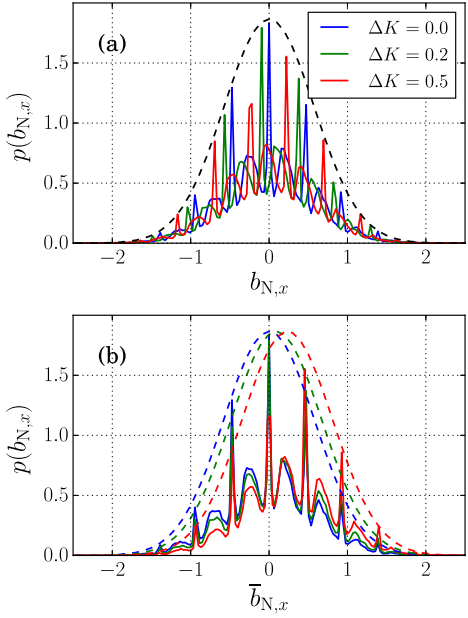}
\caption{(a) Density distribution of the $x$ component of the
Overhauser field $b_\mathrm{N}$ for deviations from the resonance
$\Delta K$, recorded for $K=200$ after the $20\,000^\mathrm{th}$
pulse. (b) Shifted distribution $\overline{b}_{N,x}$ by $\frac{2\pi
\Delta K}{T_\mathrm{R}}$. The dashed lines: the corresponding shifted
Gaussian envelope.}

\label{fig:hist_rescond}
\end{center}
\end{figure}

$p(b_{N,x})$ is shown for different $\Delta K$ in Fig.\ \ref{fig:hist_rescond}. In all distributions,
the distance between every other peak remains constant, and the
envelope follows a Gaussian distribution with a mean value of zero and a variance of $1/3$. 
Depending on the magnitude of $\Delta K$, however, the peak positions shift  
to adjust for the two resonance conditions for the Overhauser field. 
After accommodating displacement induced by the off-resonance  external magnetic field
into the Overhauser field by plotting $p(b_{N,x})$  vs $\overline{b}_{N,x}=  b_{N,x} +2\pi \Delta K / T_\mathrm{R}$
the peak positions coincide. The peak heights, however, are asymmetric due to the shifted Gaussian envelope as illustrated
in the lower panel of Fig.\ \ref{fig:hist_rescond}.

\begin{figure}[t]
\begin{center}
\includegraphics[width=0.45\textwidth,clip]{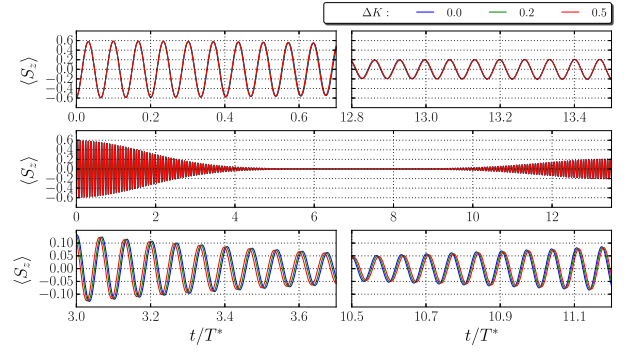}
\caption{Dynamics of the $z$ component of the central spin for
deviations from the resonance $\Delta K$ after the
$20\,000^\mathrm{th}$ pulse. Top panel: self focussing of the electron
spin immediately before and after the pulse.  Middle panel: envelope
of the ensemble average. Bottom panel: frequency shifts at
intermediate times.  }

\label{fig:CSt_rescond}
\end{center}
\end{figure}

These shifted resonance positions are understood in terms of the
resonance conditions \eqref{eqn:even_rescond} and
\eqref{eqn:arctan_rescond} where the effective Larmor frequency enters
rather then the external magnetic field.  Consequently, our
calculations back the conjectured notion \cite{Greilich28092007} of a
self focusing central spin dynamics by the dynamical redistribution of
$p(b_{N,x})$ due to the periodic laser pumping. This is illustrated in
Fig.\ \ref{fig:CSt_rescond} where the averaged electron spin response
is plotted for two different off-resonant external magnetic fields in
comparison with a resonant field.  The top panel demonstrates the
congruent dynamics immediately after and before the pulse.  Only at
intermediate times, small dephasing between the response of different
QDs are observable, as shown in the bottom two panels of Fig.\
\ref{fig:CSt_rescond}.

\subsection{Single QD vs QD ensemble}

\begin{figure}[t]
  \centering
  \begin{center}
\includegraphics[width=0.42\textwidth,clip]{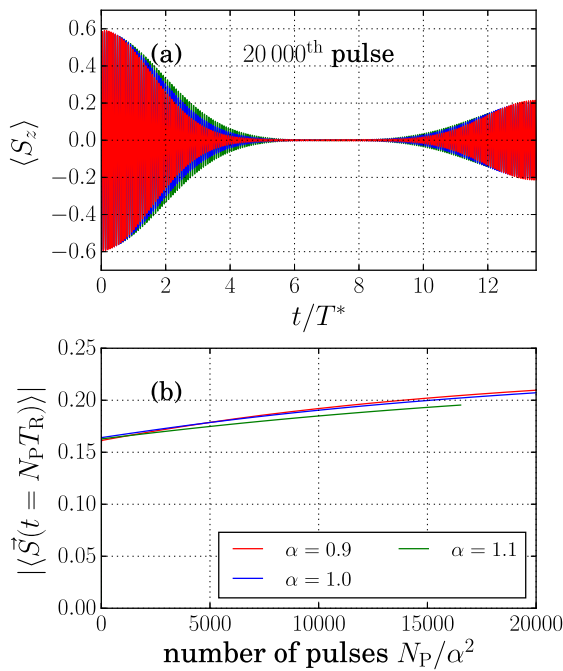}
\end{center}

\caption{Dynamics of the $z$- component of the electron spin for different scaling factors $\alpha$ of the time scale $T^*$.
(a) dynamics of $S_z(t)$ after the $20\,000^\mathrm{th}$ pulse. (b) the modulus of the revival of the electron spin
vs the  number of pulses. The external magnetic field is given by $K=200$.
}

\label{fig:CSt_scaling_Tstar}

\end{figure}

The different QDs in an ensemble not only differ in their $g$ factors
but also in their hyperfine constants $a_k$. Since it has been
established \cite{merkulov} that the key quantity for describing the
decoherence induced by the hyperfine interaction is given by $T^*$, we
parameterize the individual difference of a QD to a fictitious
reference QD characterized by $T^*$ via a scaling factor $\alpha$
\begin{align} T_\alpha^* = \alpha T^*.
\end{align} $\alpha$ depends on the different growth processes and the
distribution of radii of the QD. Here we investigate only small
variations from $\alpha = 0.9$ to $\alpha = 1.1$: larger $\alpha$
implies a slower dephasing of the central spin.  The difference in the
central spin dynamics for three different $T_\alpha^*$ is depicted in
Fig.\ \ref{fig:CSt_scaling_Tstar}(a). $T_\alpha^*$ determines the
characteristic time scale of the inital decoherence as well as of the
revival since it defines the width of the Gaussian envelope function
of the central spin dynamics.

\begin{figure}[t]
\begin{center}
\includegraphics[width=0.42\textwidth,clip]{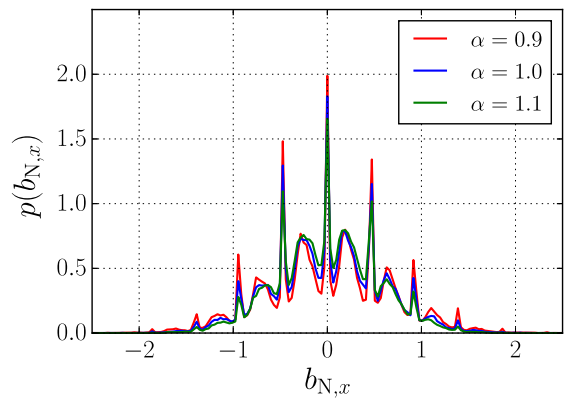} 
\caption{Density distribution of the $x$ component of the Overhauser field $b_\mathrm{N}$ after 20\,000 pulses for different scaling factors $\alpha$ of the time scale $T^*$.}
\label{fig:hist_scaling_Tstar}
\end{center}
\end{figure}

Fig.\ \ref{fig:hist_scaling_Tstar} shows that the variation of
$\alpha$ does not affect the peak positions of the distribution.
We can conclude that the sub-set of QDs resonantly pumped by the laser
pulse leads to an in-phase interference of the central spin dynamics.
Therefore, the results obtained by the simulation of a single QD help
understanding the dynamics of the whole QD ensemble.

The peak height, however, increases with decreasing $\alpha$ as
expected from the feedback mechanism of the Overhauser field and the
Knight field: the smaller $\alpha$, the larger the hyperfine coupling,
the faster the build-up of the distribution function.
Fig.\ \ref{fig:CSt_scaling_Tstar}(b) also illustrates this effect of
$T_\alpha^*$ onto the time evolution of the revival amplitude of the
central spin. Since we already discussed the influence of the number
of nuclear spins $N$ onto the time evolution, we can plot the
amplitude versus $N_P/\alpha^2$ to accommodate the leading effect of
$\alpha$. The plots demonstrate the scaling, confirming the underlying
feedback mechanisms between electron and nuclear spin system via the
Overhauser field and Knight field. However, deviations are observable
for $\alpha=1.1$.  We attribute that to the fact that the ratio
$T_\mathrm{R}/T^*_\alpha$ changes in comparison to Sec.\
\ref{subsec:numberPulses} where this ratio was kept constant.

\subsection{Influence of nuclear Zeeman effect}
\label{subsec:nuclearZeeman}


While in experiments, the nuclear Zeeman effect cannot be switched of,
we performed numerical simulations for the different ratios
$z=\frac{g_k\mu_\mathrm{N}}{g_\mathrm{e}\mu_\mathrm{B}}= 0$, $1/800$,
$1/500$. As discussed above, $z=1/800$ corresponds to the typical
experimental situation of a GaAs based QD and has been used in all
previous calculations of this paper.  $z=1/500$ is the highest
realistic ratio given by the $g$ factor for ${}^{71}\mathrm{Ga}$ with
$g_k = 1.7$.

\begin{figure}[t]
\begin{center}
\includegraphics[width=0.42\textwidth,clip]{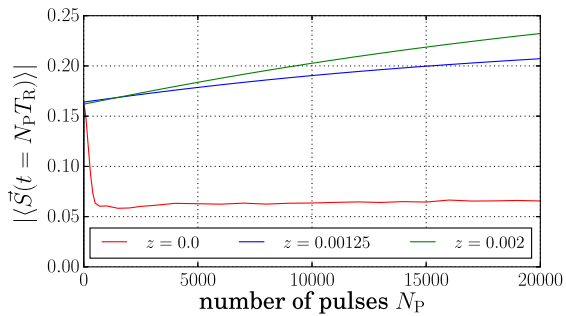} 
\caption{Evolution of the electron spin amplitude for different $z$.
The data for $z=1/800$ are taken from Fig. \ref{fig:hist_number_of_pulses}.
Parameters as in Fig.\ \ref{fig:hist_number_of_pulses}. 
}
\label{fig:revival_Zeeman}
\end{center}
\end{figure}

We found a striking difference in the revival amplitude for $z=0$ in
comparison to $z>0$ as shown in Fig.\ \ref{fig:revival_Zeeman}.  The
data for $z=1/800$ included here have already been plotted in Fig.\
\ref{fig:revival_Bext}.  While the build-up of the revival amplitude
increases slightly by artificially doubling of the nuclear Zeeman
term, the $z=0$ result shows a fundamentally different behavior.
Initially, the revival spin polarization is identical for all cases,
since it is of purely electronic origin. After some 100 pulses, the
feedback of the electron spin polarization on the nuclear spin system
becomes relevant. For $z=0$, the revival amplitude rapidly decreases
and is stabilized at a rather low value of $0.06$.

\begin{figure}[t]
\begin{center}
\includegraphics[width=0.45\textwidth,clip]{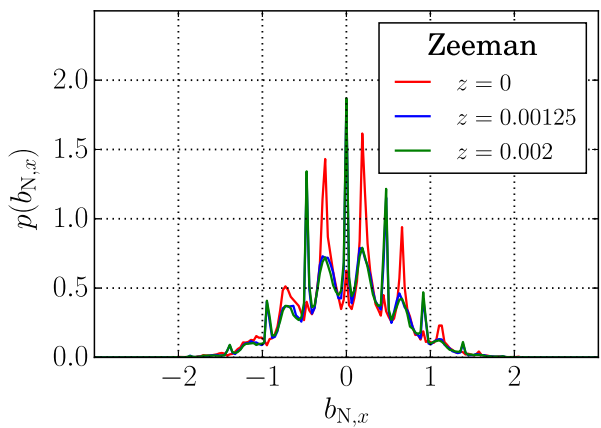} 
\caption{Density distribution of the $x$ component of the Overhauser field $b_\mathrm{N}$ for different ratios $z = \frac{g_k\mu_\mathrm{N}}{g_\mathrm{e}\mu_\mathrm{B}}$.}
\label{fig:hist_Zeeman}
\end{center}
\end{figure}

We present the corresponding Overhauser field distribution
$p(b_{N,x})$ in Fig.\ \ref{fig:hist_Zeeman}. While the shape of the
envelope remains Gaussian and the distance as well as the position of
the peaks stays the same, the weights of these peaks differ
significantly. Only marginal difference are observed for the two
finite $z$ values. For $z=0$, the weights have shifted almost
completely to the sub-set of peaks connected to the resonance
condition \eqref{eqn:arctan_rescond} corresponding to an additional
phase shift of $\Delta \omega_\mathrm{L} T_\mathrm{R} = \Delta\alpha=
2\arctan(\omega_\mathrm{L}/\overline{\gamma})$, accumulated during the
laser repetition time $T_\mathrm{R}$, comparted to the integer
resonance condition \eqref{eqn:even_rescond}.  Our findings perfectly
agree with a recent fully quantum-mechanical investigation of the mode
locking \cite{BeugelingUhrigAnders2016} in the absence of the nuclear
Zeeman effect.

\begin{figure}[t]
\begin{center}
\includegraphics[width=0.42\textwidth,clip]{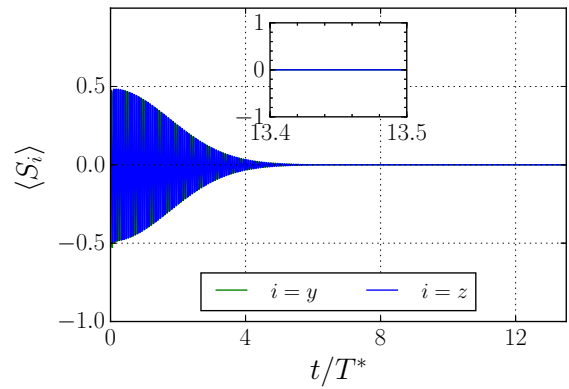} 
\caption{The toy model for a ratio of 1:3 between the peaks of the resonance sub-classes shows no electron spin revival.}
\label{fig:ToyModel_13ratio}
\end{center}
\end{figure}

In order to gain some better understanding of this surprising decay,
we used the distribution $p(b_{N,x})$ as a guide and resort to the toy
model presented in Sec.\ \ref{sec:toymodel}.  Peaks are found in
$p(b_{N,x})$ fullfiling both resonance conditions,
\eqref{eqn:even_rescond} and \eqref{eqn:arctan_rescond}. Assuming a
ratio of $1:3$ between Gaussian envelope function corresponding to the
peaks defined by \eqref{eqn:even_rescond} and respectively the peaks
defined by \eqref{eqn:arctan_rescond}, allows one to superimpose the
results for the toymodel depicted in Fig.\
\ref{fig:ToyModelRevival}(a) and \ref{fig:ToyModelRevival}(b) with
these modified spectral weights.  This leads to a finite
spin-polarization after the laser pulse which completely destructively
interferes before the next laser pulse as depicted in
Fig. \ref{fig:ToyModel_13ratio}.

Therefore, the rapid decrease of the revival amplitude for $z=0$
plotted in Fig.\ \ref{fig:revival_Zeeman} to a small finite value is
related to the strong weight imbalance between the two sub-sets of
peaks. A slightly different broadening and deviations from the trial
ratio $1:3$ are responsible for a small but finite revival amplitude.

We emphasize that the toy model phenomenologically explains the low
revival amplitude but not the deeper reason for the strong weight
imbalance between the peak heights of the two sub-classes.

It has been conjectured, that the imbalance between the peak weights
of the two resonance conditions might be attributed to the nuclear
spin precession.  In our SCA, we do not see any indication of the
reported quantum mechanical effects \cite{BeugelingUhrigAnders2017}.
No indication for a transfer of weight between both resonance
conditions when altering $z$ or $K$ have been observed in the results
obtain by our approach.

\subsection{Influence of the distribution function $p(a)$}

In this section, we extent our investigation to the influence of
different distributions $p(a)$ for the coupling constants $a_k$ on
$p(b_{N,x})$. The distribution used for the data labeled $m=2$ is
defined by Eq.\ \eqref{eqn:hackmanDist}, while 
for $m=3$  the exponential
distribution of the coupling constants is given by 
{$a_k\propto
\exp(-\lambda (k-1))$ with $k = 1..N$ and $\lambda=(r_0^3/N)=0.1$
 \cite{SchliemannKhaetskiiLoos2003,PhysRevB.76.014304, Seifert2016}, 
 see Eq.\ \eqref{eqn:expAk}.

Fig.\ \ref{fig:hist_coupling_dist} shows $p(b_{N,x})$ for these two
non-constant $p(a)$ and a fixed number of laser pulses in comparison
with the box model, where $a_k=1/\sqrt{N}$.
The distribution of the Overhauser fields still features the two
classes of peaks inside the Gaussian envelope. The differences can be
seen in the speed of the Overhauser field build-up. Distributions with
non-equal coupling constants lead to a faster development of the
Overhauser field distribution.

This corresponds to a faster build-up of the revival for non-constant
$p(a)$, see Fig.\ \ref{fig:revival_coupling_dist}. Since the coupling
constant enter quadratically into the change of the Overhauser field,
\begin{eqnarray}
\frac{d}{dt} \vec{B}_N &=& 
\sum_k a^2_k \vec{S}\times \vec{I}_k
+ z \vec{b}_{\rm ext} \times  \vec{B}_N 
\label{eq:dynamics-overhauser-field}
\end{eqnarray}
the change is dominated by the larger coupling constants for fixed
$T^*$. A non-constant distribution is therefore equivalent on a
reduced number of nuclear spins in the box model plus distribution
specific corrections, cf.\ Sec.\ \ref{scalingArg}.

\begin{figure}[t]
\begin{center}
\includegraphics[width=0.42\textwidth,clip]{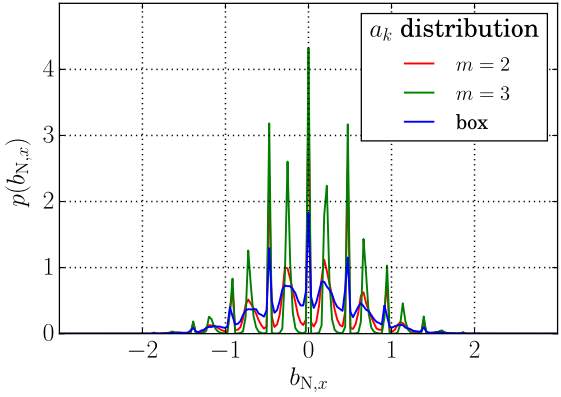} 
\caption{Density distribution of the $x$ component of the Overhauser field $b_\mathrm{N}$ after 20\,000 pulses for different distributions of the coupling constants. The external magnetic field is given by $K=200$. $N=100$ and a cut-off radius $r_0=1.5$ in \eqref{eqn:hackmanDist} for $m=2$. 
For $m=3$,  $\lambda=0.1$ in Eq.\ \eqref{eqn:expAk}, i.\ e.\ $r_0\approx 2.15$.
}
\label{fig:hist_coupling_dist}
\end{center}
\end{figure}

\begin{figure}[t]
\begin{center}
\includegraphics[width=0.4\textwidth,clip]{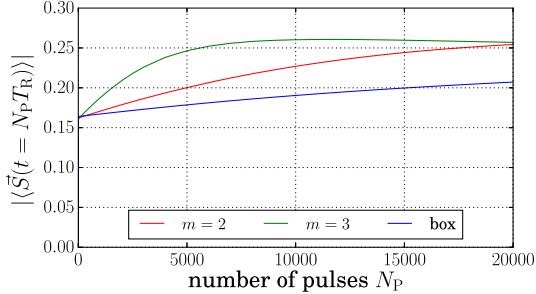} 
\caption{Revival of the central spin amplitude for different distributions of the coupling constants.
Parameters as in Fig.\ \ref{fig:hist_coupling_dist}.}
\label{fig:revival_coupling_dist}
\end{center}
\end{figure}

\subsubsection{$N$ dependent scaling behavior for distributed coupling constants}
 
%

In order for a potentially speed-up the numerics, the scaling behavior
on the number of nuclear spins for non-constant $p(a)$ is very
important. Contrary to the box model, each spin must be simulated
individually hence the run time is proportional to $N\cdot
N_\mathrm{P}$, and the validity of the scaling argument is even more
desirable.  To test its applicability the distribution given by Eq.\
\eqref{eqn:hackmanDist} with $m=2$ and $r_0=1.5$ was chosen.

%
%
\begin{figure}[t]
\begin{center}
\includegraphics[width=0.4\textwidth,clip]{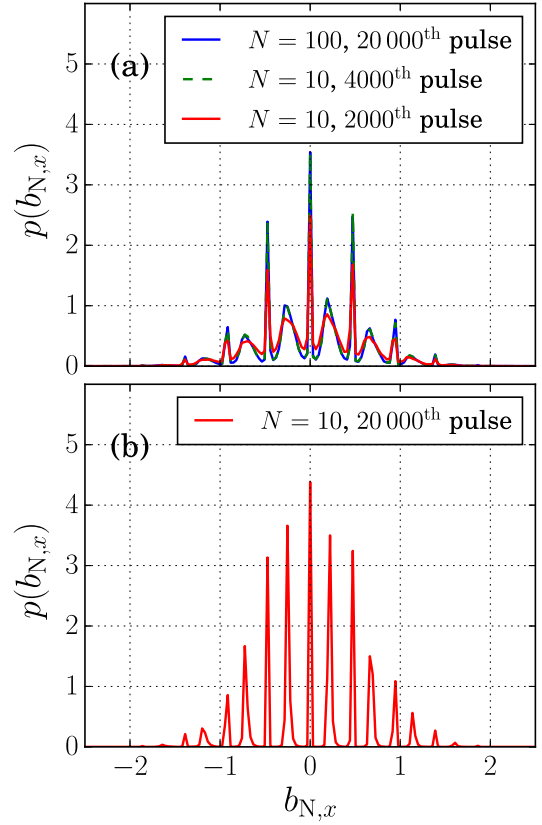} 
\caption{Scaling behavior for $N=100$ and $N=10$ with coupling constants which are distributed according to Eq.\ \eqref{eqn:hackmanDist} with $m=2$ and $r_0 = 1.5$ in an external magnetic field of $K=200$.}
\label{fig:hist_NumberOfNuclearSpins_coupling_dist}
\end{center}
\end{figure}

The results for $p(b_{N,x})$ using the same distribution function
$p(a)$ is shown in Fig.\
\ref{fig:hist_NumberOfNuclearSpins_coupling_dist} for the combinations
$(N,N_\mathrm{P})=(100, 20000), (10, 2000), (10, 4000)$.  Clear
deviations from the $r = N_\mathrm{P}/N$ scaling established only for
the box model are noticeable: $p(b_{N,x})$ for $(100, 20000)$ almost
coincides with the results for the combination $(10, 4000)$.  While in
the box model, all nuclear spins rotate synchronized, in general,
different nuclear spins have different precession speeds.

We have demonstrated that $r$ has to be replace by a distribution
dependent scaling variable $x=r f(p(a),N)$ where the deviation from
the box model scaling has to be included in the unknown correction
$f(p(a),N)$ depending on the distribution function $p(a)$ as well as
the total number of samples taken. We can estimate the ratio
$f(p(a),100)/f(p(a),10)=2$ for the single data point provided by Fig.\
\ref{fig:hist_NumberOfNuclearSpins_coupling_dist}: We need a larger
number of pulses compared to number of nuclei to achieve the same
scaling behavior exhibited in the box model. This implies that
$f(p(a), N_\mathrm{A}) < f(p(a), N_\mathrm{B})$ if
$N_\mathrm{A}<N_\mathrm{B}$.
  
This shows that computation time in the full classical model can be
reduced by a smaller system size not only because the argument
presented in \ref{scalingArg} still holds but also because, in
contrast to the box model, less nuclear spin EOM are required to be
solved.

\section{Experimental studies of the mode spectrum}
\label{sec:experimental.mode-spectrum}

\begin{figure}[t]
\begin{center}

\includegraphics[width=0.4\textwidth,clip]{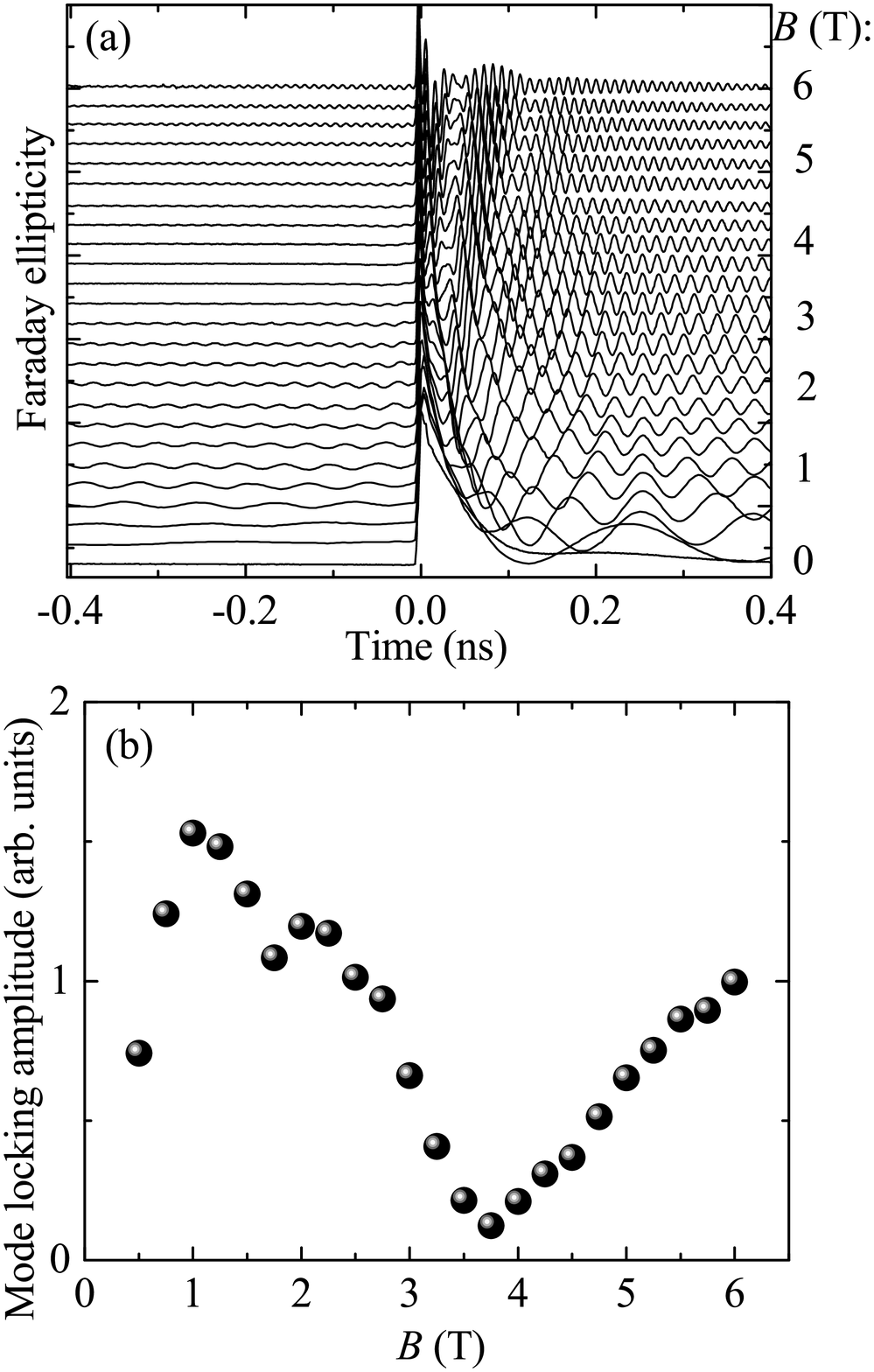}

\caption{(a): Faraday rotation measurements as function of delay between pump and probe for magnetic fields applied in the Voigt-configuration  varied between 0 and 6 T. (b) The amplitude of the mode-locked signal before the pump pulse as derived from these measurements is shown in the right panel. }
\label{fig:A1}
\end{center}
\end{figure}

\begin{figure}[t]
\begin{center}

\includegraphics[width=0.4\textwidth,clip]{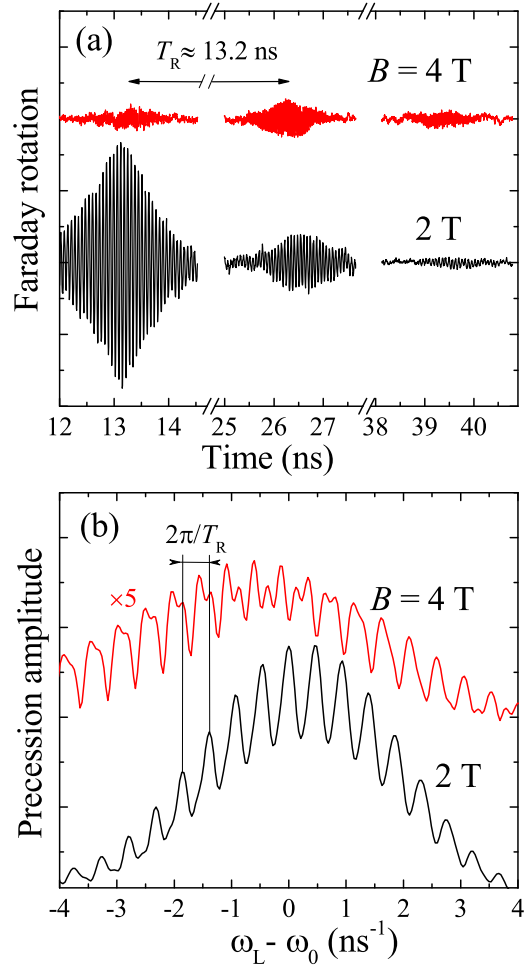}

\caption{(a): Faraday rotation measurements of the spin dynamics recorded after pumping the system by periodic pulses with 13.2 ns separation. The pumping is then switched off and the Faraday rotation is recorded over longer time scales in which periodic revivals occur due to the previous periodicity of the pumping.
(b) Taking the Fourier transform of this trace provides sufficient accuracy to resolve the precession modes. In particular at B = 4 T additional modes can be seen which are an indication for the additional modes according to Eq.\ \eqref{eqn:arctan_rescond}.}

\label{fig:A2}
\end{center}
\end{figure}


Experimental access to the precessional mode spectrum can be gained
through Faraday rotation measurements, in which the impact of the
periodic pump pulses onto the electron spins in quantum dots is traced
by a linearly polarized probe pulse whose polarization change is
measured after transmission through the sample. The spin precession
dynamics about a perpendicular magnetic field is determined by varying
the delay between pump and probe, from which the precessional mode
spectrum can be retrieved by taking the Fourier
transform. Further, by choosing a delay between pump and probe that is only slightly shorter than the pulse separation time, we can measure directly the mode-locking amplitude due to the spin revival.

Corresponding studies on InGaAs/GaAs quantum dot ensembles
so far had revealed only modes which fulfill the condition that the
precession frequency is an integer of the laser repetition rate
according to Eq.\ \eqref{eqn:even_rescond} and dominate the spectrum
(we call them integer modes in the following for brevity). Indications for modes that do not fulfill this condition but can be associated with modes fulfilling 
Eq.\ \eqref{eqn:arctan_rescond} had not been observed.

We have carefully repeated Faraday rotation studies in order to find indications for the additional modes predicted by Eq.\ \eqref{eqn:arctan_rescond}.
Details about the experiments can be found in Refs. \cite{PhysRevLett.96.227401,Greilich28092007}.
The challenge in these experiments is to scan a large enough temporal
range in time to obtain sufficient resolution in frequency space. This
is complicated by the variation of the electron g-factor in the
studied dot ensembles, which lead to a fast dephasing of the signal
and a corresponding broadening of the precession modes. Further, also a more complex form of the hyperfine coupling or additional interactions such as dipole-dipole couplings may lead to a more complex behavior of the experimental data.

As suggested by the theory, some indications for the additional modes may be found from the
amplitude of the mode-locking signal right before the next pump
pulse where the spin revival occurs. The modes that fulfill the integer spin revolution criterion Eq.\ \eqref{eqn:even_rescond}
add constructively to this amplitude. On the other hand, the modes
associated with Eq.\ \eqref{eqn:arctan_rescond} can add to the
amplitude if their frequency is not too different from the integer
modes. However, if they are located around the middle between these
modes, their orientation is opposite to the one of the modes
fulfilling Eq.\ \eqref{eqn:even_rescond}.
These modes then destructively contribute to the total amplitude of the mode-locked signal.


The interplay of these two types of modes can be varied through
varying the magnetic field amplitude. Corresponding magnetic field
measurements from B=0 up to 6 T are shown in Fig.\ \ref{fig:A1}(a),
where we focus on the amplitude of the mode-locked signal right before
the pump. Clearly the amplitude of this signal shows a non-linear
dependence with increasing magnetic field as confirmed by the magnetic
field dependence of the mode-locked signal amplitude shown in Fig.\
\ref{fig:A1}(b). This strong variation may be related to the
calculation results in Fig.\ \ref{fig:hist_Bext} which indicate that the signal amplitude
does not show a simple variation with magnetic field, but a much more
complex behavior, even though the results there are not fully converged.

From the data in Fig.\ \ref{fig:A1} one can in particular see that the
mode-locked amplitude becomes particularly weak at about 4 T. 
This may
correspond to a situation where the modes according to Eq.\
\eqref{eqn:even_rescond} and those according to Eq.\
\eqref{eqn:arctan_rescond} almost compensate each other. For other
magnetic fields the non-integer modes after Eq.\ \eqref{eqn:arctan_rescond} influence the mode-locking
amplitude apparently much weaker.

To get a more direct proof of these modes we have irradiated the quantum dot sample for an extended period of pump pulses and have switched off
then the pump, to monitor the free evolution of the spin ensemble. The
ensemble dynamics then shows revivals that occur periodically with a
separation equal to the separation between the laser pulses in the
previously applied pump protocol. To obtain sufficient resolution, we
have recorded the Faraday rotation signal over several of these echoes
as long as they show significant amplitudes.

Fig.\ \ref{fig:A2} shows a corresponding Faraday rotation trace (top
panel, recorded at 2 T and 4 T) and the corresponding Fourier
transform (bottom panel). Indeed, the spectrum at 2 T is dominated by
the integer spin revolution modes. However, at 4 T side modes appear,
whose frequencies do not fulfill the criterion of Eq.\
\eqref{eqn:even_rescond}, which have not been reported before. 
We want to highlight that these modes are prominent around the field 
strengths where the mode-locked spin amplitude shows a minimum, 
providing a consistent phenomenology.  This is a clear signature that
indeed not only the integer precession modes after Eq.\
\eqref{eqn:even_rescond} appear, but also 
additional modes contribute to the time-periodic steady state.

The goal of this experimental augment is the demonstration that the precessional mode spectrum is more complex than being just given by  
Eq.\ \eqref{eqn:even_rescond}, rather than to claim quantitative agreement with the calculations of spectral positions and amplitudes of the additional modes according to Eq.\ \eqref{eqn:arctan_rescond}.
Such agreement cannot be expected, not
only because of the ensemble study but also because of the much larger
number of nuclei of about $10^5$ in each dot in combination with a more complex distribution of hyperfine couplings.

Additional interactions such as the electric quadrupolar
interaction \cite{Sinitsyn2012,hackmannPRL2015,Glasenapp2016} as well
nuclear dipole-dipole interactions \cite{HansonSpinQdotsRMP2007}
neglected in the simulations are also expected to lead to a broadening of
the peaks in the Overhauser distributions, therefore, to a reduction
in the steady state revival amplitude. While the experiments clearly
reach the steady-state, the theoretical revival amplitude has not been
converged even after 20 000 pulses, as can be seen from Fig.\
\ref{fig:revival_Bext}. 

The experimental data presented in Fig.\ \ref{fig:A1} clearly
demonstrate a non-monotonic dependency of the mode locking amplitude
on the external magnetic field.  The discussion of the toy-model in
the Secs.\ \ref{sec:toymodel} and \ref{subsec:nuclearZeeman} suggests
that a vanishing of the mode locking amplitude might originate in the
different  amplitude  ratios for the even and the odd resonance revival 
contributions.

\section{Summary and conclusion}
\label{sec:conclusion}

We have derived a semi-classical description of the system, also
encompassing the trion decay, for the simulation of a periodically
pulsed QD. Using the FOA, we derived two classes of steady state
resonance conditions: one depends only on the repetition rate of the
pulse, $\omega_\mathrm{L} T_\mathrm{R} = 2\pi n$ and the other is also
influenced by the trion decay rate via $\omega_\mathrm{L} T_\mathrm{R}
= 2\arctan(\omega_\mathrm{L}/\gamma) + 2\pi n$.  
By the means of a simple toy
model, we have analytically shown how the Overhauser field
distribution and the electron spin dynamics, especially the revival of
the electron spin immediately before the next pulse, are connected in
the limit of large external magnetic fields.

Nuclear self-focussing was demonstrated in the
build-up of the Overhauser field distribution as well as in the
revival of the central spin signal employing 
the full semi-classical simulation of the model for equal
coupling constants.
The theoretical predictions of the
peak positions also hold  for non-constant Overhauser fields with only
a small margin of error. For large external magnetic fields
$b_\mathrm{ext}(K>100)$ the peaks are placed at integer multiples of
$\pi$ and the electron spin revival increases over time while the
behavior for smaller external magnetic fields exhibits peaks shifted
by the arctan and an electron spin revival decrease with an increasing
number of pulses.

It has been shown that larger numbers of pulses are accessible in the
box model at the same computational effort by reducing the system size
and exploiting the scaling properties defined by the variable $r =
N_\mathrm{P}/N$.
This scaling argument is used to 
make conjectures about the steady state for
realistic numbers of nuclear spins after several seconds of pulsing
that is not directly accessible to our numerical simulations.

We have investigated the QD ensemble features by including the effects
of $g$-factor variations as well as the change of the characteristic
time scale $T^*$ from QD to QD.  We have demonstrated that the
electron spin dynamics shortly after and shortly before each pulse is
essentially independent of the individual properties of each QD, and
the steady-state is determined by a Floquet condition. The Overhauser
field distribution displayed self-focussing by shifting the peak
positions to accommodate the resonance conditions.
This was reflected by the congruent central spin dynamics
immediately before and after the pulse.  The different hyperfine
coupling constants in each QD lead to a rescaling of the
characteristic time scale $T^*$. Larger hyperfine couplings do not
only cause a shorter dephasing time but also induce
a faster build-up of
the Overhauser field distribution and the electron spin revival. At
the end, the Floquet condition imposes the self-focussing
superposition of the dynamics of different QDs and a congruent central
spin behavior. Therefore the investigation of the dynamics in a single
QD can be used to gain an understanding of the ensemble properties.

The different isotopes of the QD are modeled by the   
ratio $z$ between the nuclear Zeeman energy
and the electron  Zeeman  energy. 
While realistic, non-zero values of $z$ lead to a
similar behavior in the Overhauser field and the central spin
dynamics, $z=0$ stands out. The peaks of the class of odd resonance
condition are pronounced and only a minuscule electron spin revival is
observed similar to what has been reported for a fully quantum
mechanical treatment of the problem for a small number of nuclei
\cite{BeugelingUhrigAnders2016}.

For non-equal coupling constants the computation time increases
drastically since all EOM for each individual spins have to be solved
in order to achieve reliable results for the long-time asymptotic. The
basic features  such as the position of the Overhauser peaks or the
increase of electron spin revival remain untouched.
The build-up speed of the
peak amplitudes, however, as well as the spin revival 
amplitude is different for different distributions of hyperfine fine couplings
for the same the number of pulses.
While a reduced number of nuclear spin still leads to a faster
convergence to the steady state the scaling behavior is not as
pronounced as it is for the box model: The reduction of the number of
nuclei in the simulation is less efficient.

%
%

One of the main findings of the calculations is the claim of the existence of additional precession modes besides those described by 
Eq.\ \eqref{eqn:even_rescond}. Only those had been reported in experimental studies so far. By designing an experiments with proper resolution in frequency space we could indeed resolve additional modes which may be related with those fulfilling Eq.\ \eqref{eqn:arctan_rescond}. 
These modes should lead to a reduction of the spin revival, which has been confirmed for the magnetic field strengths where they appear most prominently in the spectra. On the other hand, at field strengths where they hardly are observable the spin mode-locking amplitude is large. It will be an effort for future activities to provide a quantitative comparison of experimental data with model calculations. This will require elaborating tools (spectroscopy on refined samples) by which the precession spectra can be measured with even higher resolution in combination with calculations which are extended towards the steady state and in which further relevant interaction are included.


\begin{acknowledgments}

We are very thankful for fruitful discussions on the project with
W. Beugeling, B. Fauseweh, G\"otz Uhrig and  M. Glazov.
We acknowledge the financial support by the
Deutsche Forschungsgemeinschaft and the Russian Foundation of Basic
Research through the transregio TRR 160.
\end{acknowledgments}

\appendix

\section{Unitary transformation of density operator via an ideal $\pi$ laser pulse}\label{app:A}

The density operator of the electronic subsystem including the
trion is transformed according to 
\begin{eqnarray}
\rho^\mathrm{ap} &= & \hat{T} \rho^\mathrm{bp} \hat{T}^\dagger
\end{eqnarray}
where $\hat{T}$ is a unitary operator accounting for the laser pulse.
Under resonance conditions one finds \eqref{eqn:unitary-pulse-operator}
\begin{eqnarray}
\hat{T}&=& \mathrm{i} \ket{\uparrow\downarrow\Uparrow}\bra{\uparrow} + \mathrm{i}\ket{\uparrow}\bra{\uparrow\downarrow\Uparrow} + \ket{\downarrow}\bra{\downarrow}
\end{eqnarray}
for an ideal $\pi$-pulse. Starting from the initial density matrix 
\begin{eqnarray}
\rho^\mathrm{bp} &=&
\left(
\begin{array}{ccc}
 \rho_{\uparrow\uparrow}  &  \rho_{\uparrow\downarrow}  &  \rho_{\uparrow T}   \\
 \rho_{\downarrow\uparrow}  &   \rho_{\downarrow\downarrow} &   \rho_{\downarrow T} \\
\rho_{T   \uparrow} &   \rho_{T   \downarrow} &   \rho_{T  T} 
\end{array}
\right)
\end{eqnarray}
we arrive at
\begin{eqnarray}
\rho^\mathrm{ap} &=&
\left(
\begin{array}{ccc}
  \rho_{T  T}  &  \mathrm{i} \rho_{T   \downarrow} &  \rho_{T   \uparrow} \\
 -\mathrm{i}  \rho_{\downarrow T}   &   \rho_{\downarrow\downarrow}   & -\mathrm{i} \rho_{\downarrow\uparrow}  \\
 \rho_{\uparrow T}   &  \mathrm{i} \rho_{\uparrow\downarrow} & \rho_{\uparrow\uparrow}
\end{array}
\right)
\end{eqnarray}
Assuming that the trion was completely decayed, this matrix reduces to
\begin{eqnarray}
\rho^\mathrm{ap} &=&
\left(
\begin{array}{ccc}
0  & 0 &  0 \\
0  &   \rho_{\downarrow\downarrow}   & -\mathrm{i} \rho_{\downarrow\uparrow}  \\
0  &  \mathrm{i} \rho_{\uparrow\downarrow} & \rho_{\uparrow\uparrow}
\end{array}
\right)
\nonumber\\
&=&
\left(
\begin{array}{ccc}
0  & 0 &  0 \\
0  &  \frac{1}{2} - S_z    &  S_y -\mathrm{i}S_x  \\
0  &   S_y+\mathrm{i}S_x  & S_z + \frac{1}{2}
\end{array}
\right)
\end{eqnarray}
so that the initial electron spin is away aligned in $z$-direction after the pulse
$\vec{S}(0)= 1/2((S_z-1/2)) \vec{e}_z$.

\section{Interim results for analytical steady state solution}\label{app:B}

$S^\mathrm{bp}_z$ can be derived from the steady state condition $S^\mathrm{bp}_z = S_z(T_\mathrm{R}) $
\begin{align}
S^\mathrm{bp}_z = \dfrac{1}{2A} \left(\overline{\gamma}\omega \sin(\omega T_\mathrm{R}) - \omega^2 \cos(\omega T_\mathrm{R})\right)
\end{align}
with 
\begin{align}
A = (\omega^2+\overline{\gamma}^2) (2-\cos (\omega T_\mathrm{R})) - \overline{\gamma}\omega \sin(\omega T_\mathrm{R}) - \overline{\gamma}^2 \cos(\omega T_\mathrm{R}).
\end{align}
Then the $z$ component of the time averaged central spin is
\begin{align}
\langle S_z \rangle_{T_\mathrm{R}} = \dfrac{1}{2 A T_\mathrm{R}}\left(\overline{\gamma} (1-\cos(\omega T_\mathrm{R})) - \omega \sin(\omega T_\mathrm{R})\right).
\end{align}

%

\end{document}